\documentclass{article}
\usepackage[utf8]{inputenc}
\usepackage{amsmath}
\usepackage{amsthm}
\usepackage{amsfonts,amssymb}
\usepackage{graphicx}
\usepackage{lscape}
\usepackage{rotating}
\usepackage{longtable}
\usepackage{extarrows}
\usepackage{float}
\usepackage{hyperref}
\usepackage{mathrsfs}
\hypersetup{
    colorlinks=true, linkcolor=black,citecolor=blue}
\usepackage{geometry}
\usepackage{authblk}
\usepackage{verbatim}
\usepackage{mathtools}
\usepackage{tikz-feynman}
\usepackage{tikz}
\usepackage{slashed}
\usepackage{xcolor}
\usepackage{subfigure}
\usepackage{enumerate}

\newtheorem{thm}{Theorem}

\newcommand{\ket}[1]{|{#1}\rangle}

\DeclareMathOperator{\Tr}{Tr}
\newcommand{\tm}{t_{\text{mem}}}

\title{Large $N$ Matrix Quantum Mechanics as a Quantum Memory}
\author[1,2]{ChunJun Cao}
\author[3]{Gong Cheng}
\author[4]{Brian Swingle}
\affil[1]{Joint center for Quantum information and Computer science, University of Maryland, College Park, MD 20740, USA}
\affil[2]{Institute for Quantum Information and Matter, California Institute of Technology, Pasadena, CA
91125, USA}
\affil[3]{Maryland Center for Fundamental Physics, University of Maryland, College Park, MD 20740, USA}
\affil[4]{Brandeis University, Waltham, MA 02453, USA}

\date{}

\begin{document}
\maketitle
\begin{abstract}
   In this paper, we explore the possibility of building a quantum memory that is robust to thermal noise using large $N$ matrix quantum mechanics models. First, we investigate the gauged $SU(N)$ matrix harmonic oscillator and different ways to encode quantum information in it. By calculating the mutual information between the system and a reference which purifies the encoded information, we identify a transition temperature, $T_c$, below which the encoded quantum information is protected from thermal noise for a memory time scaling as $N^2$. Conversely, for temperatures higher than $T_c$, the information is quickly destroyed by thermal noise. Second, we relax the requirement of gauge invariance and study a matrix harmonic oscillator model with only global symmetry. Finally, we further relax even the symmetry requirement and propose a model that consists of a large number $N^2$ of qubits, with interactions derived from an approximate $SU(N)$ symmetry. In both ungauged models, we find that the effects of gauging can be mimicked using an energy penalty to give a similar result for the memory time. The final qubit model also has the potential to be realized in the laboratory.
   
\end{abstract}
\tableofcontents

\section{Introduction}

Quantum information is typically fragile when encountered in familiar physical systems, thus giving rise to the intuition that quantum information is inherently hard to preserve at macroscopic scales, e.g. in large systems or at long times. The discovery of quantum error correction showed that this intuition is incomplete: in the right kind of physical system, quantum information can be retained for an arbitrarily long time. Maintaining the information is typically an active process where we keep a vigil for errors and periodically correct them. An error correcting code operating in this mode forms an active quantum memory. Some codes may have the additional capability to preserve the encoded information for a long time naturally, without any active  intervention. A code operating in this mode is called a passive or self-correcting quantum memory---it is like a quantum hard drive. In this paper and a forthcoming companion paper, we search for passive quantum memories in the context of quantum gauge theories and models inspired by them.

Why gauge theories? For one thing, some of the most important currently known codes can be viewed as gauge theories. For example, the toric code can be viewed as a $Z_2$ lattice gauge theory with dynamical electric charges in the extreme deconfined limit. More generally, deconfined lattice discrete gauge theories are associated with a high degree of quantum entanglement and topological order, both of which are linked with quantum error correction~\cite{toriccode,KITAEV20032,KitaevPreskill}. One feature of topological order is a robustness of the ground space to perturbations of the Hamiltonian, which is suggestive of the potential for passive error correction. And while the two-dimensional toric code does not form a passive memory~\cite{Alicki2009_2d}, a four-dimensional analog does~\cite{Alicki2009_4d}. Moreover, phases of matter exhibiting topological order associated with a discrete non-Abelian group can be used for fault tolerant quantum computation, a fact that also hints at a degree of intrinsic protection~\cite{Kitaev2006}.

From another perspective, quantum error correction has been linked with quantum gravity in the context of holographic duality, also known as the AdS/CFT correspondence~\cite{Almheiri:2014lwa}. The best understood examples of the correspondence involve continuous non-Abelian gauge theories with large gauge groups, thus providing another connection between non-Abelian gauge symmetry and quantum error correction. In this context, one also finds that the boundary conformal field theory with $SU(N)$ gauge symmetry can possess a low temperature confined phase which is qualitatively distinct from the generic high temperature physics in the deconfined phase. This boundary confinement-deconfinement phase transition is dual to the Hawking-Page transition~\cite{hp} between a thermal AdS geometry and a AdS black hole geometry in the bulk \cite{Witten:1998zw}. It was suggested in \cite{thresholdadscft} that when treating the low energy sector of AdS/CFT as an approximate quantum error-correcting code under the influence of thermal errors, the encoded information living in the bulk thermal AdS geometry can be disturbed by the scattering of thermal particles with probability $O(1/N^2)$. However, in the high temperature phase, the encoded information is scrambled by the black hole, thus inducing a large logical error. In the large $N$ limit, the distinct phases correspond to the existence of a threshold, such that logical information remains robust in the low temperature phase where the errors occur less frequently\footnote{Note that unlike the tensor network constructions \cite{HaPPY} where one changes the system size by increasing the number of layers, which roughly corresponds to changing the IR cutoff in the bulk or the UV cutoff on the boundary theory when it is put on a lattice, the size of the physical system in \cite{thresholdadscft} is altered by $N$ of the $SU(N)$ symmetry while keeping the number of ``boundary sites'' fixed. Heuristically, the physical Hilbert space can be taken as placing $O(N^2)$ qudits on each boundary site.}.Therefore, it is natural to explore whether this type of robustness also extends to other theories with such gauge symmetries.

With these motivations, we have undertaken an investigation of quantum error correction in non-Abelian gauge theories with large gauge groups. We focus here on $SU(N)$ groups with $N$ taken large. This large $N$ limit is like a thermodynamic limit, and we expect that passive error correction can only emerge in such a thermodynamic limit. We define and study quantum codes in a variety of models, including models inspired by gauge theories but where we relax the gauge constraint. We are seeking to answer two key questions: (1) are there potentially useful lab-constructible quantum memories built from gauge theories or derived models and (2) can naturally occurring systems, e.g. QCD or a hypothetical holographic dual to quantum gravity in our universe, naturally preserve quantum information for a long time? This paper focuses on quantum mechanical systems with large gauge groups, while the companion paper \cite{fieldpaper} will consider quantum field theories with large gauge groups.

The key results of the present paper are as follows. First, we discuss broadly how a property called large $N$ factorization is related to an approximate version of the Knill-Laflamme error correction condition \cite{KnillLaflamme,approxKL}. Assuming a code space with the right properties can be defined and coupled to a thermal bath satisfying reasonable assumptions, we then prove a theorem establishing the existence of a memory lifetime which scales polynomially with $N$. We then consider a sequence of models of increasing generality and show how the conditions of the main theorem can be satisfied. For a certain fully gauged model, we find that the conditions of the theorem are immediately satisfied; for models where we do not gauge the symmetry or even make the symmetry approximate, an energy penalty scaling like $\log N$ is needed to meet the conditions of the main theorem. While here we focus on the construction of such quantum memories and their memory times, there are other aspects of the model that are relevant for practical implementation, such as whether quantum information information can be efficiently encoded, recovered, and processed. We will leave these discussions for future work.

In the remainder of the introduction, we first review the basic notions for readers without the necessary background in quantum codes or quantum gauge theories. We then give a high level overview of our results. We also note that a few prior works have considered the role of large $N$ factorization and/or gauge symmetry in defining quantum error correcting codes~\cite{Mintun:2015qda,Milekhin2021}. There is also a large literature on passive quantum memories from a variety of approaches~\cite{Dennisetal,Alicki2009_4d,KITAEV20032,Bacon2006}. See also \cite{Brown_review} for a more comprehensive review.

\subsection{Background on quantum codes}

In this subsection, we review the key concepts of quantum error correction needed for our discussion. This section is aimed at readers with a more high energy background.

Mathematically, a quantum error correcting code is a subspace, the \textit{logical space} $\mathcal{C}$, of a larger \textit{physical Hilbert space} $\mathcal{H}$. Errors are modeled by a quantum process $\mathcal{E}$ and error correction is given by another process $\mathcal{R}$ such that 
\begin{equation}
    \mathcal{R}(\mathcal{E}(\rho)) = \rho \quad \forall \rho \in \mathcal{C}.
\end{equation}
If the error channel is viewed as probabilistically applying errors from a set $\{E_a\}$, then one way to formulate the necessary condition for the existence of an error correction process is the Knill-Laflamme (KL) condition,
\begin{equation}
    \langle i | E_a^\dagger E_b | j \rangle = c_{ab} \delta_{ij},
\end{equation}
where $|i\rangle$ and $|j\rangle$ are any two elements of a basis for $\cal{C}$. The intuition behind KL is that errors are unable to cause transitions between distinct states in the code. We can also relax these exact equalities to obtain instead an approximate quantum code.

In practice, such a code is instantiated within a concrete physical substrate, with $\mathcal{H}$ being some model of the physical space of states of the substrate. For example, a code ($\mathcal{C}$) might be implemented as a subspace of the full energy levels ($\mathcal{H}$) of a superconducting circuit (substrate). In some cases, the code space might correspond to the ground space of some Hamiltonian for the substrate, but this is not required in general.

A code is used as follows. Information is first encoded using some encoding operation. This operation might correspond to a quantum circuit which takes a fixed initial state in $\mathcal{H}$ and produces the desired code state in $\mathcal{C}$. After encoding, the code is subject to errors, corresponding to one application of the error channel $\mathcal{E}$. Then the recovery procedure $\mathcal{R}$ is applied to recover the initially encoded information. The condition above guarantees success provided the encoding prepares a state in $\mathcal{C}$. Finally, the information can be read out by a decoding operation. Note the distinction between recovery, which repairs the information but keeps it encoded, and decoding, which reverses the encoding map when applied to an error free state\footnote{Note that somewhat confusingly, decoding in the context of active error correction can also refer to applying some suitably chosen operations to the state based on syndrome measurements. This process typically leave the information encoded, which corresponds to our recovery operation in our current definition. }. 

To make these ideas more concrete, we can consider the simple goal of information storage for a period of time (as opposed to any active computation on the encoded data). We think of the code during the error part of the protocol as sitting idle in the presence of an undesirable environment. The effect of the environment is what produces errors in the code, so the channel $\mathcal{E}$ implicitly contains information about the nature of the environment and how long the code is exposed to the environment. We denote such an error channel by $\mathcal{E}_t$, where $t$ corresponds to the time for which the code is exposed to the environment. Correspondingly, the recovery operation $\mathcal{R}$ may depend on $t$ and indeed on the whole structure of $\mathcal{E}_t$. 

In this paper we consider the environment to be a thermal reservoir at some temperature $T$. There is still freedom in determining exactly how the reservoir couples to the system, but from an effective theory perspective we should expect all allowed couplings to be present with some strength. In this context, the need for active error correction can be phrased like this. Given a code $\mathcal{C}$, the code requires active error correction if, as $t \rightarrow \infty$, there is no $\mathcal{R}$ that can correct $\mathcal{E}_t$. In other words, a code requires active error correction if, given enough time, the error channel will eventually degrade the information beyond the ability of any physically conceivable recovery operation to repair.

More generally, we can define a memory time $\tm$ as the time beyond which $\mathcal{E}_t$ degrades the information beyond any hope of recovery. Of course, this time depends on the code and the error model, but it also depends on what recovery operations we can conceivably perform and on how well we want to be able to recover the information (e.g. the degree of approximation in our approximate code). A perfect passive memory is one for which $\tm = \infty$. Typically, we can only hope for this sort of behavior in a thermodynamic limit, so we should really view a passive memory as a sequence of codes indexed by the physical size $n$ for which $\lim_{n\rightarrow \infty} \tm(n) = \infty$. Even more practically, what really matters is how $\tm$ scales with $n$ and other parameters of the code.

For the purpose of potentially implementing a quantum memory in the lab, we may also want to impose some additional rules of the game. One variant of such rules are called the Caltech rules \cite{Landon-Cardinal2015,Brell2016}. For example, we might obtain a long memory time by sending the energy cost of errors to infinity ``by hand'', but this could be considered ``cheating'' unless there is a good physical reason for this large energy penalty. The idea is that we indeed want to effectively forbid errors, but this should happen naturally without sending any microscopic energy scale to infinity. Of course, if one does have a system where errors are particularly costly, then it might form a good enough passive memory in the sense that $\tm$ is long enough for practical purposes. To give another example, to be able to construct codes in physical three-dimensional space, we might want some restrictions on the kinds of codes we consider, e.g. trying to avoid the need for too many long-range interactions~\cite{Brell2016}. We note that all the models we consider here do have non-local interactions between the large $N$ degrees of freedom\footnote{Note that it is somewhat ambiguous what constitutes the local degrees of freedom in the large $N$ gauged oscillator model as they are not simply tensor product of qubits. We show in later sections that there is a construction where two such gauged oscillators are spatially local with respect to each other. However, each one contains a large $N$ degrees of freedom even though there is no sense in which they occupy an extended region in space. It is in this sense that these degrees of freedom are related to each other non-locally.}.

\subsection{Background on quantum gauge theories}

In this subsection, we review the key concepts of quantum gauge theories needed for our discussion. This section is aimed at readers with a more quantum information background.

At the most basic level, a gauge theory is defined by giving a gauge group $\cal{G}$ and a set of degrees of freedom transforming under $\cal{G}$ with the requirement that all physical states and operators are invariant under $\cal{G}$. This invariance condition is known as Gauss' law, in honor of its roots in electromagnetism. An immediate question is why one would use such a redundant description at all? Experience shows that one can always use a gauge-invariant description of the states and operators, but such a description often hides important underlying features of the physics. For example, loop-like operators known as Wilson lines provide a class of gauge invariant observables in quantum field theories with gauge fields, but the extended nature of these operators can obscure aspects of locality in the theory. In special cases, one can even have an equivalence between a local gauge theory and another local theory with no explicit gauge structure, for example, the duality between the 3d statistical Ising model and 3d $Z_2$ gauge theory. 

We study such gauge theories in large part because they form a crucial part of our description of nature. The quantum theory of electromagnetism is a gauge theory where the gauge group is related to the abelian Lie group $U(1)$. The more general Standard Model is also a gauge theory with a non-Abelian gauge group of the form $SU(3)\times SU(2) \times U(1)$. Gauge theories also arise in condensed matter systems, for example, when describing so-called spin liquid phases, and, as we mentioned above, they also have interesting error correction properties. 

Apart from the motivations from known models of high-energy particle physics and condensed matter systems, another motivation to consider gauge theories comes from quantum gravity. Since the discovery of AdS/CFT, it is generally believed that certain $d$ dimensional $SU(N)$ gauge theories with large $N$ admit a dual geometric description in terms of a $d+1$ dimensional theory of gravity. From a quantum information perspective, this correspondence provides the possibility of hiding quantum information deep inside the emergent geometry. This has inspired a lot of research  in the field of  holographic quantum error correcting codes~\cite{Almheiri:2014lwa}, with explicit models such as the HaPPY code~\cite{HaPPY}, the effective field theory code~\cite{Furuya:2021lgx}, and many other related constructions using tensor networks \cite{Harris:2018jfl,Cao:2020ksw,Dolev:2021ofc}. Current holographic codes often utilize the extra dimension, such that any local/UV errors can not access the information deep inside along the extra radial direction. However, very little of the error correcting power in these constructions come directly from the gauge symmetries in AdS/CFT. In this work, we will focus on a different kind of protection from the large $N$ gauge symmetries, which can be related to the small Newton's constant in these holographic theories  \cite{Milekhin2021,thresholdadscft}. This opens another possibility to design a new kind of quantum memory.  

An interesting structure we exploit in the large $N$ gauge theory is its sparse density of excitations at low energy. Due to Gauss' law, the only allowed excitations of a $SU(N)$ gauge theory on any compact manifold are gluons that combine into gauge singlets. Therefore, the density of states is determined by counting the number of gauge singlets in a given energy level. It is a general feature that the density of states in such model scales exponentially with some power of energy and does not depend on $N$. As is pointed out in \cite{Aharony:2003sx}, even for a free gauge theory with large $N$, the projection to the gauge invariant subspace induces effective interactions among the gluons, which is negligible at high energy but dominates at the low energy regime. In fact, for a weakly coupled $SU(N)$ gauge theory, there is a phase transition analogous to the comfinement/deconfinement transition in QCD, in the limit of $N\rightarrow \infty$. As we show, this induces a critical temperature, $T_c$, below which the quantum information can be long-lived but above which the information is quickly destroyed. In the geometric description, this is believed to be dual to the Hawking-Page transition and the high temperature phase can be thought of as a black hole that kills the quantum information~\cite{thresholdadscft}. 

To zero in on just the effects of gauge symmetry, we distill only the necessary ingredients from the above gauge theory intuitions and consider a gauge theory with no spatial geometry at all. It turns out that even the simplest $0+1$-dimensional gauge theory of harmonic oscillators is sufficient for our purposes. The inverted harmonic oscillator has been studied as a dual to a two dimensional string theory and can be considered as toy model of holography~~\cite{Itzhaki:2004te,Berenstein_2004,Klebanov:1991qa,Johnson:2021tnl}. Here we start from a similar but non-inverted harmonic oscillator model and try to build a quantum memory from it. 

\subsection{Organization and overview of results}
The rest of this paper is organized as follows. In Section~\ref{setup}, we describe in detail the large $N$ matrix quantum mechanics model with $SU(N)$ gauge symmetry. We then discuss how it may be construed as an approximate error correcting code and define various bath models we will use for the rest of the paper. We then discuss the mutual information diagnostic for memory time and distill the core conclusions of our work in the form of Theorem~\ref{thm}. It shows that if a large $N$ system has a sparse low energy spectrum, couples to the bath uniformly, and is an approximate QECC, then it has a memory time polynomial in $N$ at low enough temperatures. In the ensuing Sections, we deploy this theorem to various quantum memory constructions.

In Section~\ref{example}, we study two concrete examples of quantum memories built from two gauged harmonic oscillator models coupled to a thermal reservoir. The memory time scales as $N^2$ in both models. For the first model, the $N^2$ life time persists even at high temperature, but the model requires non-local logical operations. We then construct a local model which has a $N^2$ memory time in the low temperature phase, but requires a large coupling of strength $\sim \log N$ between the gauged oscillators. At zero temperature limit $T\rightarrow 0$, the first model has a constant $N^2$ life time, while the second model has a divergent life time that scales as $N^2e^{\frac{\omega}{T}}$, where $\omega$ is some energy scale of this model. 

In Section~\ref{case3}, 
we consider what happens if the gauge constraint is not imposed exactly, but only as an energetic cost in the Hamiltonian. A large number of non-singlet excitations now modify the physics, but the information retention time can be still be quadratic in $N$ if the energy scale of non-singlet excitations is large. 

Finally, in Section~\ref{spin}, we ask to what extent does robustness depend on the symmetry being exact and the Hilbert space of quantum systems being infinite dimensional. On the practical level, we want the quantum memories to be implementable on finite dimensional quantum systems like qubits or qudits. To this end, we consider an even more relaxed situation where the $SU(N)$ symmetry itself is not only ungauged but also approximate. We then build a quantum memory consisting of $N^2$ number of  physical qudits that interact non-locally with each other. Qualitatively, this model behaves similarly to the one in Section~\ref{case3} in the low energy sector, except small correction terms that suppressed to order at most $\sim \log N/N$. We find that if we require the total energy of the interaction scales as $\log{N}$ and a reasonable pattern of coupling strengths, then Theorem 1 again holds to guarantee a $N^2$ memory time. The zero temperature limit of life times for the last two models also diverge as $N^2e^{\frac{\omega}{T}}$. 

\section{Matrix model and set up}
\label{setup}

\subsection{Matrix Quantum Mechanics}

We wish to study a gauged matrix model of  harmonic oscillators. First, recall the classical Lagrangian of a single harmonic oscillator,
\begin{equation}
L=   \frac{1}{2} (\dot{x}^2-\omega^2 x^2).
\end{equation}
The canonical momentum is $p = \frac{\partial L}{\partial \dot{x}} = \dot{x}$, in terms of which we also have the Hamiltonian description, 
\begin{equation}
    H = \frac{1}{2} (p^2 + \omega^2 x^2).
\end{equation}
The system is quantized in the usual way by setting $[x,p]=i\hbar$.

A generalization can be made by promoting the variable $x(t)$ to an $N\times N$ matrix $X(t)$ and promoting products of variables to matrix multiplication. We hasten to emphasize that this $N \times N$ matrix structure does not refer to a quantum Hilbert space; rather it is a way of organizing $N^2$ variables into a single unit. Then one gets an action for a matrix harmonic oscillator,
\begin{equation}\label{matrixlagrangian}
    L=\frac{1}{2}\Tr[(\partial_t X)^2-\omega^2X^2].
\end{equation}
Again, let us emphasize that the trace is not a quantum Hilbert space trace but rather a sum over the matrix degrees of freedom. The diagonal elements of $X$ are real variables, while the off-diagonal components are complex variables coming in conjugate pairs. 

The momentum conjugate to $X^i_j$ is $P^j_i=\dot{X}^j_i$. We quantize the theory by specifying canonical commutators for all pairs of conjugate variables. If we extend quantum Hermitian conjugation to include a tranpose of the matrix degrees of freedom, then the quantum variable $X$ is Hermitian. In terms of its matrix elements $X^i_j$, Hermiticity implies that $X^i_i$ are quantum Hermitian while $X^i_j$ and $X^j_i$ are quantum Hermitian conjugates, $(X^i_j)^\dagger = X^j_i$. Note that $X^i_j$ is not quantum Hermitian when $i \neq j$, but the above formula is still correct. 

After quantization, it is convenient to work in terms of creation and annihilation operators, 
\begin{equation}
\begin{split}
    a^i_j&=\sqrt{\frac{\omega}{2}}(X^i_j+iP^i_j)\\
    a^{\dagger j}_i&=\frac{1}{\sqrt{2\omega}}(X^j_i-iP^j_i),
\end{split}
\end{equation}
which satisfy
\begin{equation}\label{commutation}
    [a^i_j,a^{\dagger k}_l]=\delta^i_l\delta^k_j.
\end{equation}
The Hamiltonian written in terms of $a$ and $a^{\dagger}$'s takes the simple form \begin{equation}\label{Hamiltonian}
    H=\omega\Tr(a^{\dagger}a)
\end{equation}
after subtracting the zero-point energy. 
There is a unique vacuum state $|0\rangle$ satisfying 
\begin{equation}
    a^{i}_j|0\rangle =0, ~\forall i,j.
\end{equation}
Acting creation operators on this vacumm then generates all the Fock states that span the Hilbert space. 

Looking back at the  Lagrangian in Eq.~\eqref{matrixlagrangian}, we see that it has a symmetry under unitary transformations  $X(t)\rightarrow UX(t)U^{\dagger}$, with $U\in U(N)$. (This is why we consider complex off-diagonal entries in $X$.) This can be made into a local symmetry $U(t)$, meaning a symmetry where $U(t)$ can be different at each different time, as follows. We add a matrix gauge field $A(t)$ and promote the time derivative to a covariant derivative, defined as 
\begin{equation}
D_tX =\partial_t X- [A,X]
\end{equation}
with $A(t)$ transforming as $A(t)\rightarrow U(t)A(t)U(t)^{\dagger}-i U(t)\partial_tU(t)^{\dagger}$.  One can check that the covariant derivative transforms in the same way as $X(t)$, which is $D_tX\rightarrow U(t)D_tXU(t)^{\dagger}$. Therefore, the Lagrangian is invariant under this transformation. We call this symmetry with respect to the time-dependent transformation the gauge symmetry. 

The model equipped with such a symmetry can be regarded as a redundant description of a physical system whose Hilbert space is defined to include only gauge-invariant states. Indeed, in this model, $A(t)$ can be view as a Lagrange multiplier. Fixing a gauge $A=0$, and integrating out $A$ in the path integral representation gives the Gauss law constraint:
\begin{equation}
    G=[X,\dot{X}]=0.
\end{equation}
This condition is implemented by requiring that physical states $|\psi\rangle$ be annihilated by $G$, $G|\psi\rangle=0$.

One can see that the states satisfying this constraint are the symmetry singlets, generated by traces of operators acting on the vacuum, 
\begin{equation}\label{eqstates}
   \ket\psi=\Tr(a^{\dagger n_1})\Tr(a^{\dagger n_2})\cdots \Tr(a^{\dagger n_k})\ket 0.
\end{equation}

These states, after being normalized, span a subspace of the ungauged matrix model's Hilbert space. They are also eigenstates of the Hamiltonian (\ref{Hamiltonian}) such that
\begin{equation}
    H|\psi\rangle=\omega\sum_{i=1}^k n_i\ket \psi.
\end{equation}

\subsection{Physical states and the Knill-Laflamme Condition}

We now describe in broad terms how the properties of these gauge-singlet states can be leveraged to generate approximate codes obeying an approximate version of the KL condition. First some terminology. If a gauge invariant operator contains only a single trace, it is called single trace operator. Operators with multiple traces are called multi-trace. Using the commutation relation, Eq.~\eqref{commutation}, one can derive a useful commutator rule for single trace operators, with $n_{1,2}$, $m_{1,2}$ being $O(1)$ number,
\begin{equation}
    [\Tr(a^{n_1}a^{\dagger m_1}),\Tr(a^{n_2}a^{\dagger m_2})]=(n_1m_2-n_2m_1)\Tr(a^{n_1+n_2-1}a^{\dagger m_1+m_2-1})+O\left(\frac{1}{N^2}\right).
\end{equation}

Using these rules, overlaps of states generated by acting multi-trace operators on the ground state can be obtained to leading order in large $N$. For example, the squared norm of the state $\Tr(a^\dagger) | 0\rangle$ is 
\begin{equation}\label{eqnorm}
    \langle0|\Tr(a^n)\Tr(a^{\dagger n})|0\rangle=\langle 0|[\Tr(a^n),\Tr(a^{\dagger n})]|0\rangle=nN^n+O\left(\frac{1}{N^2}\right).
\end{equation}
We also find that states with the same energy are almost orthogonal, with overlaps suppressed by a power of $N$. For example, 
\begin{equation}
    \langle 0|\frac{\Tr(a^m)\Tr(a^{n-m})}{N^{\frac{m}{2}}}\frac{\Tr(a^{\dagger n})}{N^{\frac{n}{2}}}|0\rangle=O\left(\frac{1}{N}\right).
\end{equation}

Moreover, one can show that a factorization property holds for products of single trace operators. As an example, given four single trace operators with vanishing expectation value, $O_i$, $O_j$, $O_k$ and $O_l$, we have
\begin{equation}\label{eqfact}
\begin{split}
    \langle 0| O_iO_kO_l^{\dagger}O_j^{\dagger}|0 \rangle &\sim \langle 0|O_iO_j^{\dagger}|0\rangle \langle 0|O_kO_l^{\dagger}|0\rangle +\langle 0|O_iO_l^{\dagger}|0\rangle \langle 0|O_kO_j^{\dagger}|0\rangle +O\left(\frac{1}{N^2}\right)\\
    \langle0|O_iO_kO_j^{\dagger}|0\rangle&\sim O\left(\frac{1}{N}\right).
\end{split}
\end{equation}
If we could find a way to classify $O_i$ and $O_j$ as logical operators and $O_k$ and $O_l$ as errors, then this equation would be interpreted as an approximate version of Knill-Laflamme condition (aKL). While one may deviate from KL in many forms, the approximate KL condition we consider in this work could be written as 
\begin{equation}\label{eqknill}
\begin{split}
    \langle i|E_a^{\dagger}E_b|j\rangle&=f_{ab}\delta_{ij}+\frac{g^{ij}_{ab}}{N^2}\\
    \langle i|E_a|j\rangle&=\frac{e_a^{ij}}{N}.
\end{split}
\end{equation}

In the above formulae, we are imagining that $E_a$ and $E_b$ are some general non-identity error operators. In general, we expect them to be any gauge invariant operators that satisfy some constraints, depending on how we couple the model to a thermal environment. We will give concrete examples in Section~\ref{example}, and explicitly calculate the functions $f$, $g$ and $e$. These examples will show such large $N$ matrix quantum mechanics models can used to construct approximate quantum error correcting codes~\cite{approxKL}.




For later use, we also briefly discuss the counting of gauge invariant operators. If we have only one matrix harmonic oscillator, the general invariant states are in the form of Eq.~\eqref{eqstates}. At energy level $n\omega$, the number of distinct states (which are also approximately orthogonal) is equal to the number of partitions of integer $p(n)$. For example, at level $n=3$. The three partitions are, 
\begin{equation}
    \begin{split}
   & 3=3+0\\
   & 3=2+1\\
   & 3=1+1+1.
    \end{split}
\end{equation} 
Correspondingly, we have three gauge invariant states at energy $3\omega$, which are
\begin{equation}
    \begin{split}
        \Tr(a^{\dagger 3})|0\rangle, \ \ \Tr(a^{\dagger 2})\Tr(a^{\dagger})|0\rangle, \ \ \Tr(a^{\dagger})^3|0\rangle.
    \end{split}
\end{equation}

In general, we can have arbitrary $k\leq n$ number of matrix harmonic oscillators, each carrying energy that is an integer multiple of $\omega$. We label the individual matrix operator by its subindex, $a_1$, $a_2$, $a_3$, $\cdots$. The counting becomes more complicated with multiple oscillators, so let us focus on the states generated by single trace operators. At energy level $n\omega$, these states can be written as 
\begin{equation}\label{eqphstate}
    \Tr(P\{a_1^{\dagger n_1}a_2^{\dagger n_2}\cdots a_k^{\dagger n_k}\})|0\rangle,
\end{equation}
with $n=\sum_i n_i$. The state $|0\rangle$ is tensor product of $k$ ground states of each oscillator.  We use the symbol $P$ to denote a particular permutation of all the operators inside the bracket. By counting how many  different ways of splitting $n$ and the number of permutations, one can show that the number of these single trace operators asymptotically scales as $k^n$ when $n$ is large enough. For simplicity, in the concrete models that will be discussed in the following sections, we consider no more than two matrix modes.

\subsection{Bath model}\label{bathmodel}

We have seen that if errors and logical operators can be chosen properly, then an approximate Knill-Laflamme condition holds. Based on this, we expect that a single error operator can only corrupt the logical information at order $\frac{1}{N^2}$. At the same time, gauge invariance constrains the number of possible errors that may occur in the form of excitations. Since the density of states for the gauged oscillator grows with energy but is independent of $N$ below some critical value, we may expect such a system to serve as a good quantum memory even at finite temperature. To show this explicitly, we couple the proposed quantum memory to a thermal reservoir and calculate the memory time $\tm$.  

We make the standard assumption that the bath is Markovian and couples to the system locally. More specifically, we model the bath as collection of bosonic modes $b_l$, e.g. a collection of oscillators, with the following thermal spectral function~\cite{PhysRevA.82.022305},
\begin{equation}\label{eqthermalfactor}
    \Tr(\rho_B b_k(\nu)b_l^{\dagger}(-\nu)) =\delta_{kl} \left|\frac{\nu}{1-e^{-\beta\nu}}\right|:=\delta_{kl} \gamma(-\nu).
\end{equation}
Physically, this means that the probability of having a given error that adds energy $\epsilon=-\nu$ to the system is suppressed by a thermal factor $\gamma(\epsilon)$. 

We will discuss three scenarios for the structure of the bath. In particular, there is a degree of ambiguity in what constitutes a generic physical bath for these systems, so we consider a number of models of varying levels of generality.

Case 1: Consider two matrix harmonic oscillators, $a_1$ and $a_2$, separated by a large distance in space. The reason for introducing two such oscillators will become clear in the next section. We model this setup by assuming that these oscillators couple to the thermal bath independently with order one coupling constants. This means we assume that thermal errors can include any multi-trace operators provide they are uncorrelated between $a_1$ and $a_2$. By uncorrelated error, we mean that the modes $a_1$ and $a_2$ can not appear inside the same trace; instead errors must be of the form 
\begin{equation}
    E_{L,R}=\prod_{(r_1, r_2)\in L, \ (s_1, s_2)\in R}\frac{ :\Tr(P\{a_1^{\dagger r_1}a_1^{r_2}\})::\Tr(P\{a_2^{\dagger s_1}a_2^{s_2}\}):}{N^{\frac{r_1+r_2+s_1+s_2}{2}}}
\end{equation}
with $L$ and $R$ being some sets of integer 2-tuples. The symbol $P$ denotes the sum over all distinct arrangements of the  creation and annihilation operators inside the bracket. The symbol $::$ means
normal ordering and the $N$-dependent normalization is to ensure that the single trace operator of
each $a$, $a^{\dagger}$ arrangement has order one norm (see Eq.~\eqref{eqnorm}). In terms of these errors, the thermal Hamiltonian is 
\begin{equation}
    H_{\rm thermal}=\sum_{L,R}\lambda_{L,R} b_{L,R} E_{L,R}+ h.c.,
\end{equation}
where h.c. stands for hermitian conjugate. We will discuss this case in more detail in Section~\ref{case2}. 

Case 2: We assume the coupling with the bath modes $b_l$ are through single-trace operators, with order one coupling constants $\lambda_l$'s. We still consider modes $a_1$ and $a_2$ from two different matrix oscillators. The thermal coupling Hamiltonian is in the form of
\begin{equation}
    H_{\rm thermal}=\sum_{\{n_k\}} \lambda_{\{n_k\}} b_{\{n_k\}} \frac{:\Tr(P\{a_1^{\dagger n_1}a_2^{\dagger n_2}a_1^{n_3}a_2^{n_4}\}):}{N^{\frac{\sum_{k=1}^4 n_k}{2}}}+h.c., 
\end{equation} 
At first sight, it appears that coupling only to single-trace operators is a strong assumption, because in general we should allow all gauge invariant operators to couple with the bath. However, this is still general enough since single-trace errors accumulate into multi-trace errors as time increases. We will discuss this scenario in more detail in Section~\ref{case1}.

Case 3:  We consider the most general coupling to the thermal bath where each of the $N^2$ harmonic oscillator modes can couple to the bath independently without requiring the coupling terms be gauge invariant. In addition, we assume that the coupling constant decays exponentially with the number of creation and annihilation operators involved. This is a reasonable assumption because the correlated errors that involves more operators occur with smaller probability.  We explain this in detail in Section~\ref{case3}, where we consider the ungauged model.

\subsection{Mutual information diagnostic}\label{MIdiag}

Given the assumptions on the bath described above, the time evolution of our system can be described by a Lindblad equation after tracing out the bath,
\begin{equation}
    \dot{\rho}(t)=-i[H,\rho(t)]+ \sum_a \lambda_a(\nu_a) E_a\rho(t)E_a^{\dagger}-\frac{1}{2}\sum_a \lambda_a(\nu_a) \{E_a^{\dagger}E_a,\rho(t)\},
\end{equation}
where $\lambda_a(\nu_a)=\lambda_a\gamma(\nu_a)$, is the coupling constant multiplied with thermal factor. $\nu_a$ is the energy of  excitation, and the $\{E_a\}$ depend on the interaction Hamiltonian. 

The generic expectation is that the above system dynamics will lead to rapid information loss within the system. We wish to show that this is not the case for the right kind of code constructed from the matrix degrees of freedom. To quantitatively measure the memory time, we entangle the encoded $d$-dimensional system with a reference $R$. The initial state is thus
\begin{equation}
\begin{split}
    &\frac{1}{\sqrt{d}}\sum_{i=1}^d|i\tilde{i}\rangle :=\frac{1}{\sqrt{d}}\sum_{i=1}^d|i\rangle _S|i\rangle _R\\
    &\rho_0=\frac{1}{d}\sum_{i,j=1}^d|i\tilde{i}\rangle \langle j\tilde{j}|,
    \end{split}
\end{equation}
and the correlation between the reference is maximal.

The goal is then to understand how the system-reference correlation evolves with time. The state evolves under the Lindblad dynamics into 
\begin{equation}\label{expansion}
\begin{split}
    \rho(t)&=e^{\mathcal{L} t} \rho_0,\\
    \textrm{where}~~\mathcal{L} \mathcal{O}&=\sum_k \lambda_kE_k\mathcal{O}E_k^{\dagger}-\sum_{k}\frac{\lambda_k}{2}\{E_k^{\dagger}E_k,\mathcal{O}\} .
\end{split}
\end{equation} 
In terms of the system and reference reduced density matrices $\rho_S(t)=\Tr_R[\rho(t)]$ and $\rho_R(t)=\Tr_S[\rho(t)]$, we may quantify the correlation using the mutual information, 
\begin{equation}\label{MI}
    I(S:R)(t)=S(\rho_S(t))+S(\rho_R(t))-S(\rho(t)) .
\end{equation}
When the mutual information is maximal, $I = 2 \ln d$, we say that the quantum information is stored in the quantum memory. However, as $I(S:R)$ decays, the information of the encoded qubit leaves the memory and dissipates into the environment. For the sake of simplicity, we will focus on quantum memory constructions that store a single encoded qubit in the ensuing sections, so $d=2$.

We now tie together all the ingredients above in a theorem. In general, different couplings with the bath can induce different errors on the system. However, as long as such errors are not too numerous and can be suppressed by the code, i.e., satisfying the aKL condition, then the following theorem guarantees the memory time of the quantum information is relatively long. To be precise, this theorem shows that the mutual information diagnostic decays polynomially with $t$ at a rate suppressed by $\frac{1}{N^2}$, which implies that the memory time can scale polynomially with $N$.

\begin{thm}\label{thm}
Consider a system whose low energy states form an approximate quantum error correcting code that encodes a logical qudit $S$ and a set of physical errors $\mathcal{E}$ induced by the system-bath coupling (c.f. Section~\ref{bathmodel}). Provided the following conditions are satisfied, 
\begin{enumerate}
    \item (Sparse spectrum) There exists an energy $\epsilon_0$ below which the effective number of errors\footnote{These are errors that contribute non-trivially in the Lindblad equation.} in $\mathcal{E}$ with energy $\epsilon\leq \epsilon_0$ is independent of $N$ and bounded by $\exp(\mu\epsilon)$ for some $\mu>0$,
    \item (Uniform coupling) The thermal Hamiltonian that couples each $E\in\mathcal{E}$ to an independent bath operator has coupling constant no larger than $O(1)$,
    \item\label{cond3} (Approximate error correction) $\forall E_a, E_b\in \mathcal{E}$, the approximate Knill-Laflamme condition (Eq.~\eqref{eqknill}) is satisfied,
\end{enumerate}
then the mutual information $I(S,R)$ defined in Eq.~\eqref{MI} is given by
\begin{equation}\label{eqmain}
I(S:R)(t)=2\ln{d}-K\left(\frac{t}{N^2}\right),
\end{equation}
where $K(x)$ is a polynomial function of $t$, for temperature $T<\frac{1}{\mu}$.

\begin{proof}
The proof is given in Appendix~\ref{apentropy}.
\end{proof}
\end{thm}

In the following sections, we will construct different systems coupled via different bath models, as outlined in Section~\ref{bathmodel}, and prove that they satisfy the assumptions in the theorem, such as the approximate Knill-Laflamme condition  (aKL)  for a corresponding set of error operators and logical states.

\textbf{Remark: }
Note that the above theorem does not require the code subspace to be the ground space of the system. Therefore, the $N^2$ scaling of memory time holds generally for these models no matter how the code subspace is chosen as long as the requisite conditions are satisfied. Intuitively, this is because the transition rate between states are $1/N^2$ suppressed. However, as excited states can still slowly decay into lower energy states at the same rate, their memory times need not be longer by lowering the temperature. 

On the other hand, for quantum memories where we also choose the code subspace to be the ground space, the memory lifetime can be further extended by a multiplicative factor $\exp(\omega/T)$, which is related to the effect of thermal suppression. This renders the total low temperature memory time to be of $O(N^2\exp(\omega/T))$ for some energy scale $\omega$ as we see in some of the examples below.

 \section{Fully gauged model }\label{example}
 \subsection{Non-local model}\label{case2}
 
As the first example, consider two gauged harmonic oscillators $a_1$ and $a_2$ with $SU(N)$ gauge symmetry, separated by a large distance.  Their interaction is  weak, and each couples to the thermal reservoir locally and independently\footnote{More realistically, thermal noise can have finite correlation. In such cases, we assume that the distance separating the two oscillators is much larger than the correlation length so our assumptions remain good approximations.}. The Hamiltonian takes the following form, 

\begin{equation}\label{eqfull}
    H=\omega \Tr(a_1^{\dagger}a_1)+\omega \Tr(a_2^{\dagger}a_2)+\sum_{L,R}\lambda_{L,R}b_{L,R}E_{L,R}+h.c. ,
\end{equation}
where $E_{L,R}$'s are the possible errors acting on this model, and $\lambda_{L,R}$'s are  $O(1)$ coupling constants.  

The general gauge invariant states are given in Eq.~\eqref{eqphstate} with $k=2$. However, since we assume that these modes only couple to the thermal bath locally, the set of possible error operators is a proper subset of all physical operators. These errors are uncorrelated in the sense that each error operator contains only a single trace either the mode $a_1$ or the mode $a_2$, but never both.  Given two sets of integer 2-tuples $L$ and $R$, we can define the errors as, 
\begin{equation}\label{eqerror}
    E_{L,R}=:\prod_{(r_1, r_2)\in L, \ (s_1, s_2)\in R}\frac{ \Tr(P\{a_1^{\dagger r_1}a_1^{r_2}\})\Tr(P\{a_2^{\dagger s_1}a_2^{s_2}\})}{N^{\frac{r_1+r_2+s_1+s_2}{2}}}:. 
\end{equation}
where the symbol $P$ denotes the sum over all distinct arrangements of the $a_i,a_i^{\dagger}$ operators\footnote{For example, $a^{\dagger 2}a^2$ and $a^{\dagger}aa^{\dagger}a$ are counted as two distinct arrangements. }. In general, $P\{a^{\dagger r_1}a^{r_2}\}$ admits $(r_1+r_2)!/r_1!r_2!$ terms of distinct arrangements. Each operator is brought to normal ordering, and divided by some power of $N$ to make sure that these operators have $O(1)$ norm.

\subsubsection*{Logical states}

Given this set of error operators, we can construct logical states that are protected against them. In general, we may choose the logical states to be
 
 \begin{equation}
    |\mathcal{I}\rangle :=\prod_{(m,n)\in \mathcal{I} } \frac{\Tr(a_1^{\dagger m}a_2^{\dagger n})}{N^{\frac{m+n}{2}}}|0\rangle_{12} .
\end{equation}
with $m,n\geq1$ and belonging to some set $\mathcal{I}$ of integer 2-tuples. The state $|0\rangle_{12}$ is the tensor product of the vacuum states of the two oscillators. For simplicity, let us focus on a code subspace spanned by two states and treat it as a logical qubit\footnote{The logical subspace we use is not the lowest energy subspace of the Hamiltonian, so they have the tendency to decay into lower energy states. However, the decay rate is suppressed by $\frac{1}{N^2}$, as discussed in Appendix~\ref{apphase}.}:
\begin{equation}\label{eqnlls1}
\begin{split}
    |\tilde{1}\rangle&=\frac{\Tr(a_1^{\dagger 2}a_2^{\dagger 2})}{N^2}|0\rangle_{12}\\
    |\tilde{2}\rangle&=\frac{\Tr(a_1^{\dagger}a_2^{\dagger})^2}{\sqrt{2}N^2}|0\rangle_{12} .
\end{split}
\end{equation}

They are not orthogonal states, since $\langle\tilde{1}|\tilde{2}\rangle=\frac{\sqrt{2}}{N}+O(\frac{1}{N^3})$. However, it is straightforward to identify an orthonormal basis $|\tilde{\uparrow}\rangle$ and $|\tilde{\downarrow}\rangle$, defined as
\begin{equation}\label{eqnlls2}
\begin{split}
    &|\tilde{\uparrow}\rangle=|\tilde{1}\rangle\\
    &|\tilde{\downarrow}\rangle=\frac{|\tilde{2}\rangle-\langle\tilde{2}|\tilde{1}\rangle|\tilde{1}\rangle}{\sqrt{1-\langle\tilde{2}|\tilde{1}\rangle^2}}.
\end{split}
\end{equation}

\subsubsection*{Errors and aKL}

Because we assume errors are gauge invariant, they are expressible as single or multi-trace operators (see Eq.~\eqref{eqerror}). For the purpose of invoking Theorem~\ref{thm}, it suffices to check that aKL holds for these errors,

\begin{equation}
\begin{split}
    \langle\tilde{i}|E_{L',R'}^{\dagger}E_{L,R}|\tilde{j}\rangle&=f_{L'R',LR}\delta_{ij}+\frac{g_{L'R',LR}^{ij}}{N^2}+O(\frac{1}{N^4})\\
    \langle\tilde{i}|E_{L,R}|\tilde{j}\rangle&=\frac{e_{L'R',LR}^{ij}}{N}+O(\frac{1}{N^3}).
\end{split}
    \label{eqn:aKLnonlocal}
\end{equation}

To build up some intuition, let us first examine a few instructive examples before we arrive at the more general relation \eqref{eqn:aKLnonlocal}. To start, one can check that the operator $\frac{\Tr(a_1^{\dagger}a_1)}{N}$ has the logical states as eigenstates,
\begin{equation}
    \frac{\Tr(a_1^{\dagger}a_1)}{N}|\tilde{i}\rangle=\frac{2}{N}|\tilde{i}\rangle.
\end{equation}
Therefore, it is a logical identity operator. We can similarly check that the Hamiltonian that generates the system dynamics is proportional to the logical identity, which is necessary for a memory. 

Next, let us examine a class of single-trace errors of the form $E_{n}=\frac{\Tr(a_1^{\dagger n})}{\sqrt{n}N^{\frac{n}{2}}}$. As a potential bit flip error, we check the transition amplitude,
\begin{equation}\label{eqh}
    \langle \tilde\downarrow|E_{m}^{\dagger}E_{n}|\tilde\uparrow\rangle =O(\frac{1}{N^3}).
\end{equation}
The bit flip error only carries $n$ dependence at order $O(\frac{1}{N^3})$. Similarly, the phase errors are also suppressed,
\begin{equation}\label{phase}
    \begin{split}
        &\langle \tilde\uparrow|E_{m}^{\dagger}E_{n}|\tilde\uparrow\rangle =\delta_{nm}\left[1+\frac{3n}{N^2}+O\left(\frac{1}{N^4}\right)\right]\\
       & \langle \tilde\downarrow|E_{m}^{\dagger}E_{n}|\tilde\downarrow\rangle =\delta_{nm}\left[1+\frac{2n}{N^2}+O\left(\frac{1}{N^4}\right)\right].
    \end{split}
\end{equation}

\begin{figure}[H]
    \centering
    \begin{tikzpicture}[scale=2]
    \draw (0,0) node[below,font=\scriptsize]{$\Tr(a_1^{\dagger 2}a_2^{\dagger 2})$}--(0,1) node[above,font=\scriptsize]{$\Tr(a_1^{\dagger 2}a_2^{\dagger 2})$};
    \draw (1,0) node[below]{$E_n$}--(1,1) node[above]{$E_n$};
    \draw (2,0.5) node{$+$};
    \draw (3,0) node[below,font=\scriptsize]{$\Tr(a_1^{\dagger 2}a_2^{\dagger 2})$}-- (4,1) node[above]{$E_n$};
    \draw (4,0) node[below]{$E_n$}-- (3,1) node[above,font=\scriptsize]{$\Tr(a_1^{\dagger 2}a_2^{\dagger 2})$};
    \filldraw (3.5,0.5) circle (1pt);
    \end{tikzpicture}
    \caption{The inner product can be viewed as scattering between two particles. $(1,0)$ represents the logical state $|\tilde{\uparrow}\rangle$. The connected part is suppressed by $\frac{1}{N^2}$.}
    \label{fig1}
\end{figure}
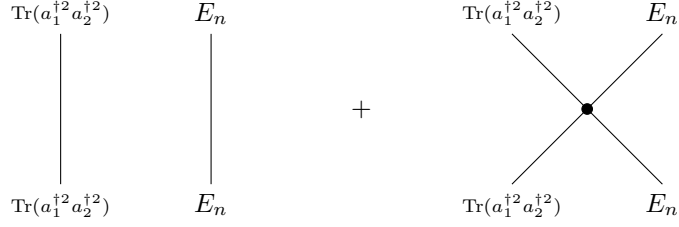

Note that the leading order terms of these equations come from factorized two point functions  $\langle 0|E^{\dagger}_{m}E_{n}|0\rangle \langle\tilde{i}|\tilde{j}\rangle$, and they represent the non-connected contribution to the four point correlator. The sub-leading corrections are from the connected part, shown diagrammatically in Figure~\ref{fig1}. It is this part that induces phase and bit-flip errors between the two logical states. In a similar way, one can check that the aKL condition holds for other single-trace errors. We present the more comprehensive calculations in Appendix~\ref{apphase}. 

\begin{figure}[H]
    \centering
   \begin{tikzpicture}
    \draw (0,0) node{$F(m,n,p,q)=$};
    \draw (2,-1) node[below]{$\uparrow/\downarrow$}--(2,1) node[above]{$\uparrow/\downarrow$};
    \draw (3,-1) node[below,font=\small]{$E_n$}--(3,1) node[above,font=\small]{$E_p$};
    \draw (4,-1) node[below,font=\small]{$E_m$}--(4,1) node[above,font=\small]{$E_q$};
    \draw (5,0) node{$+$};
    \draw (6,-1) node[below]{$\uparrow/\downarrow$}--(6,1) node[above]{$\uparrow/\downarrow$};
    \draw (7,-1) node[below,font=\small]{$E_n$} -- (7.4,-0.2) ;
    \draw (7.6,0.2)-- (8,1) node[above,font=\small]{$E_q$};
    \draw (8,-1) node[below,font=\small]{$E_m$}--(7,1) node[above,font=\small]{$E_p$};
    \draw (9,0) node{$+$};
    \draw (10,-1) node[below]{$\uparrow/\downarrow$}--(10,1)node[above]{$\uparrow/\downarrow$};
    \draw (11,-1)node[below,font=\small]{$E_n$}--(12,1) node[above,font=\small]{$E_q$};
    \draw (12,-1) node[below,font=\small]{$E_m$}--(11,1) node[above,font=\small]{$E_p$};
    \filldraw (11.5,0) circle (2pt);
   \end{tikzpicture}
    \caption{The first two diagrams are totally disconnected. The third is partly connected and represents the $O(\frac{1}{N^2})$ term in $F(m,n,p,q)$.}
    \label{fig2}
\end{figure}
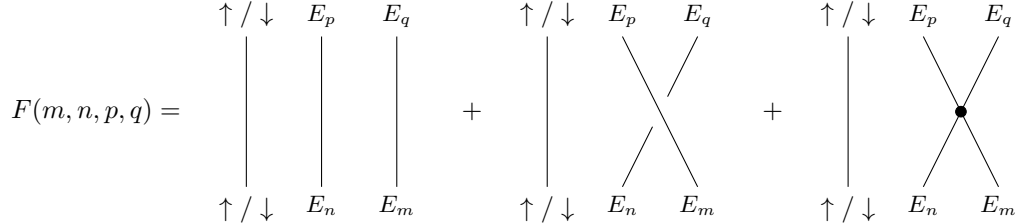

The conclusion of this analysis is that single-trace operators can only produce $1/N$ suppressed errors, thus giving rise to the approximate Knill-Laflamme condition (aKL). Using the results for single-trace errors, one can also derive the aKL conditions for any multi-trace errors. As an example,
\begin{equation}
\begin{split}
    &\langle \tilde\uparrow|E_{p}^{\dagger}E_{q}^{\dagger}E_{m}E_{n}|\tilde\uparrow\rangle =F(m,n,p,q)+\frac{G_{\uparrow}(m,n,p,q)}{N^2}+O(\frac{1}{N^4})\\
    &\langle \tilde\downarrow|E_{p}^{\dagger}E_{q}^{\dagger}E_{m}E_{n}|\tilde\downarrow\rangle =F(m,n,p,q)+\frac{G_{\downarrow}(m,n,p,q)}{N^2}+O(\frac{1}{N^4}).
\end{split}
\end{equation}
The function $F$ is the same in both correlators, so it represents a correctable error. In contrast, different $G$ functions appear in the two equations. Their difference, $G_{\uparrow}-G_{\downarrow}$, represents a small uncorrectable error. They both have simple diagrammatic representation as shown in Figure~\ref{fig2} and Figure~\ref{fig3}, from which we can read off the function $F(m,n,p,q)$. It must take the form
\begin{equation}\label{eqg}
\begin{split}
&F(m,n,p,q)=(\delta_m^p\delta_n^q+\delta_m^q\delta_n^p)+\delta_{m+n}^{p+q}\frac{f(p,q,m,n)}{N^2},
\end{split}
\end{equation}
where the $O(\frac{1}{N^2})$ term comes from the partly connected diagram in Figure~\ref{fig2}. Since $F$ does not depend on the logical information and will not contribute to uncorrectable errors, it is not pertinent to the subsequent discussion. Therefore, we will not  determine $f$ explicitly. 

The function $G$ can be read off from Figure~\ref{fig3}. Using the results for single-trace errors in Eq.~\eqref{phase}, we can show that,
\begin{equation}
\begin{split}
  &G_{\uparrow}(m,n,p,q)=(\delta_m^p\delta_n^q+\delta_m^q\delta_n^p)(3n+3m)  \\
  &G_{\downarrow}(m,n,p,q)=(\delta_m^p\delta_n^q+\delta_m^q\delta_n^p)(2n+2m).
 \end{split}
\end{equation}
From this explicit form, we see that there is indeed a difference between the these functions, so the logical states are connected by multi-trace operators, again with $1/N$ suppressed amplitudes provided the indices $m,n,p,q$ are not too large.

\begin{figure}[H]
    \centering
   \begin{tikzpicture}
    \draw (0,0) node{$G_{\uparrow/\downarrow}(m,n,p,q)=$};
    \draw (2,-1) node[below]{$\uparrow/\downarrow$}--(3,1) node[above]{$E_p$};
    \draw (3,-1) node[below]{$E_n$}--(2,1) node[above]{$\uparrow/\downarrow$};
    \filldraw (2.5,0) circle (2pt);
    \draw (4,-1) node[below]{$E_m$}--(4,1) node[above]{$E_q$};
    \draw (5,0) node{$+$};
    \draw (6,-1) node[below]{$\uparrow/\downarrow$}--(6.5,0);
    \draw (6.5,0)--(8,1) node[above]{$E_q$};
    \draw (6,1) node[above]{$\uparrow/\downarrow$}--(6.5,0);
    \draw (6.5,0)--(8,-1) node[below]{$E_m$};
    \filldraw (6.5,0) circle (2pt);
    \draw (7,-1) node[below]{$E_n$} -- (7,1) node[above]{$E_p$};
    \draw (9,0) node{$+$};
    \draw (10,0) node{$p\leftrightarrow q$};
   \end{tikzpicture}
    \caption{The $G$ functions arise from connected diagrams of order $O(\frac{1}{N^2})$ that involve the state $\Tr(a_1^{\dagger})|0\rangle $. }
    \label{fig3}
\end{figure}

Now, although we only explicitly calculated a particular class of multi-trace errors, the algorithm generalizes directly to more general cases. The analog of the $G$ function always consists of connected scattering diagram of logical states and single-trace operators. In other words, the uncorrectable part of any multi-trace error can be written as a sum over contributions of the single-trace errors, due to the factorization property of large $N$ matrix models (Eq~\eqref{eqfact}). Therefore, if the aKL condition is satisfied for the single-trace operators, it will also hold for these multi-trace errors. Note that if a diagram has four legs connected, then it must be suppressed by $\frac{1}{N^2}$. In special cases we also have three-leg diagrams, where the error operators involve $\frac{\Tr(a_{1,2}^{\dagger 2})}{\sqrt{2}N}$ or $\frac{\Tr(a_{1,2}^{\dagger})}{\sqrt{N}}$. For example, the $G$ function in the correlator $ \langle \tilde{\uparrow} | \frac{\Tr(a_{1}^{2})}{\sqrt{2}N}\frac{\Tr(a_{2}^{ 2})}{\sqrt{2}N}\frac{\Tr(a_{1}^{\dagger 2})}{\sqrt{2}N}\frac{\Tr(a_{2}^{\dagger 2})}{\sqrt{2}N}|\tilde{\uparrow}\rangle $ contains the product of two three-leg diagrams, as shown in Figure~\ref{fig4}. Each of them is proportional to  $\frac{1}{N}$, so the term involving $G$ that depends on the logical state is still suppressed by $\frac{1}{N^2}$.

\begin{figure}[H]
    \centering
    \begin{tikzpicture}[scale=1.5]
    \draw (-1,0) node[left]{$ \langle \tilde{\uparrow} | \frac{\Tr(a_{1}^{2})}{\sqrt{2}N}\frac{\Tr(a_{2}^{ 2})}{\sqrt{2}N}\frac{\Tr(a_{1}^{\dagger 2})}{\sqrt{2}N}\frac{\Tr(a_{2}^{\dagger 2})}{\sqrt{2}N}|\tilde{\uparrow}\rangle $};
    \draw (-0.5,0) node[left]{$\supset$};
     \draw (0,-1)node[below]{$\uparrow$}-- (0,0);
     \draw (0,0)--(-0.5,1) node[above,font=\small]{$\Tr(a_1^{ 2})$};
     \draw (0,0)--(0.5,1)node[above,font=\small]{$\Tr(a_2^2)$};
     \filldraw (0,0) circle (1pt);
     \draw (2,1)node[above]{$\uparrow$}--(2,0);
     \draw (2,0)--(1.5,-1)node[below, font=\small]{$\Tr(a_1^{\dagger 2})$};
     \draw (2,0)--(2.5,-1)node[below,font=\small]{$\Tr(a_2^{\dagger 2})$};
     \filldraw (2,0) circle (1pt);
    \end{tikzpicture}
    \caption{Contribution of connected diagram with three legs. Each is suppressed by $\frac{1}{N}$. }
    \label{fig4}
\end{figure}

\subsubsection*{Mutual information diagnostic}

Now we couple this quantum memory to a single-qubit reference system $R$, and calculate the mutual information between them (Eq.~\eqref{MI}). At $t=0$, $\rho_0$ is a pure state,

\begin{equation}
    \rho_0=\frac{1}{2}\sum_{i,j=1}^2|i\tilde{i}\rangle _{SR}\langle j\tilde{j}|.
\end{equation}
The system and reference are maximally entangled, so the initial mutual information is $I(S:R)(0)=2\ln{2}$.

Next we count the effective number of error operators which will contribute non-trivially to logical errors under the Lindblad dynamics. Operators that contain more than two $a_1$ or $a_2$ trivially annihilate the logical states, so we only need to count the error operators with $r_2,s_2\leq 2$.   The number of gauge invariant error operators with energy $n\omega$ that satisfy this requirement is bounded by $n^2p(n)$, where $p(n)$ is the number of integer partitions of $n$. So the number of effective errors is sub-exponential asymptotically.

By construction, the bath model has uniform coupling as required by condition 2 in theorem~\ref{thm}. Together with the aKL condition above, we apply Theorem~\ref{thm} and conclude the mutual information decays in the following way,
\begin{equation}\label{eqmain}
I(S:R)(t)=2\ln{2}-K(\frac{t}{N^2}),
\end{equation}
where $K(t)$ is linear in $t$ at early time. More specifically, when $\lambda t \ll N^2$, $K(t)\sim \frac{\sum_n n\lambda_n t}{N^2}$ . It develops a dependence on higher powers of $t$ at late time $\lambda t\sim  N^2$, where $K(t)\sim \sum_p(\frac{\sum_n n \lambda_n t}{N^2})^p$.  

Details of the proof for  Eq.~\eqref{eqmain} and more rigorous results can be found in Appendix~\ref{apentropy}, but here we briefly sketch the general ideas behind the proof. The dominant errors in the early-time regime are the single trace errors. In contrast, in the late time regime, the dominant errors are multi-trace. In fact, the average size $k$ of errors, defined as the number of traces in an error operator, grows linearly with time. These errors, when acting on logical states, cause a loss of quantum information proportional to $\frac{k^p}{N^{2p}}$, leading to the $(\frac{t}{N^2})^p$ growth behavior at late time. 

Note that the rate of information loss here is captured by the sum over $\lambda_n$, which can be approximated by 
\begin{equation}
\begin{split}
    \sum_n n^q\lambda_n=&\sum_n n^qd_ne^{-n \beta \omega}\\
    \sim& \sum_n n^{q+2} p(n) e^{-n\beta \omega},
\end{split}
\end{equation}
where $q=1$, and $d_n$ is the number of distinct states at the $n$th level. In this model, $d_n\sim n^2p(n)$, which grows sub-exponentially with $n$. As the sum is always finite, it follows that the memory time scales like $\tm \sim N^2$. From the above discussion, we conclude that this model serves as a good passive quantum memory.

Now we discuss the zero temperature limit. Since the thermal fluctuation is suppressed,  the relevant error operators are only those which lower the energy when acting on logical states. In this model, they are $\frac{\Tr(a_{1,2})}{ \sqrt{N}}$ and $\frac{\Tr(a_{1,2}^2)}{N}$, whose aKL conditions are given in Appendix~\ref{apphase}. They lead to a spontaneous decay of quantum information at a constant rate of $O(\frac{1}{N^2})$ at zero temperature.  There are also operators that don't change energy of the system, but they have no contribution to the information loss since their thermal spectral functions (Eq.~\eqref{eqthermalfactor}) vanish at zero temperature limit, $lim_{\beta\rightarrow\infty}\gamma(\nu=0)=0$.  

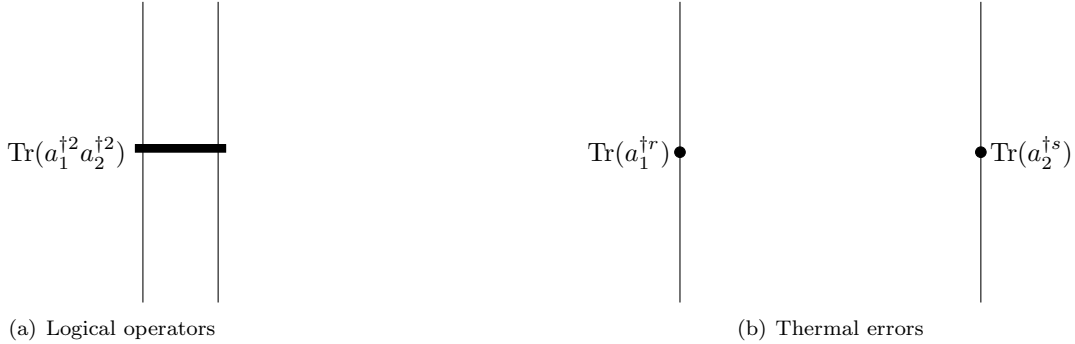
\begin{figure}[H]
\center
\subfigure[Logical operators]
{
\begin{tikzpicture}
\draw (0,-2)--(0,2);
\filldraw (-0.1,0) rectangle (1.1,0.1);
\draw 
(-0.1,0) node[left]{$\Tr(a_1^{\dagger 2}a_2^{\dagger 2})$};
\draw (1,-2)--(1,2);
\end{tikzpicture}
}
\hspace{120pt}
\subfigure[Thermal errors]
{
\begin{tikzpicture}
\draw
(7,-2)--(7,0) node[left]{$\Tr(a_1^{\dagger r})$}--(7,2);
\draw
(11,-2)--(11,0) node[right]{$\Tr(a_2^{\dagger s})$}--(11,2);
\filldraw 
(7,0) circle (2pt);
\filldraw
(11,0) circle (2pt);
\end{tikzpicture}
}

\caption{In (a), two modes are brought together to implement a logical operator. In (b), two modes are separated far away. Thermal fluctuation are sufficiently local so the quantum information is preserved. }
\label{logicalerror}
\end{figure}

To modify the encoded information, we need to apply logical operators, which involve $\Tr(a_1^{\dagger}a_2^{\dagger})$ or $\Tr(a_1^{\dagger 2}a_2^{\dagger 2})$. This requires us to perform non-local operations (or a series of local operations) by bringing the two modes together and turning on some local coupling, as shown in Figure~\ref{logicalerror}.  We assume this procedure to be accurate. After the logical operations, we can separate the two modes so that the system again behaves like a passive memory that protects the encoded information. 

We can also implement logical operations without bringing the two modes together. Instead, we may induce a coupling mediated by a third ancillary mode $c$. The procedure is that we first bring the mode $c$ close enough to the mode $a_1$ and implement the operator $\Tr(a_1^{\dagger m}c^{\dagger})$. Then we bring together the intermediate mode and the mode $a_2$ and implement the operator $\Tr(a_2^{\dagger n}c)$. These two moves can be combined together to give the desired logical operation, 
\begin{equation}
    \Tr(a_2^{\dagger 2} c)\Tr(a_1^{\dagger 2}c^{\dagger})|0\rangle \propto \Tr(a_1^{\dagger 2} a_2^{\dagger 2})|0\rangle.
\end{equation}

In summary, this model achieves the power law scaling of memory time with system size. However, it has some drawbacks. First, it requires a gauge symmetry that acts consistently on the two matrix modes separated by a large distance. Second, the logical operation is non-local, and thus can be difficult to implement. For these reasons, we construct another model, where the two matrix modes are put together, but with the added price that a high energy penalty is required.

 \subsection{Local model }\label{case1}

In this model, we still consider two matrix harmonic oscillators, but we put them together and turn on a large coupling between them. The full Hamiltonian is 
 \begin{equation}\label{eqHlocal}
    H=\omega \Tr(a_1^{\dagger}a_1)+\omega \Tr(a_2^{\dagger}a_2)+J[\frac{\Tr(a_1^{\dagger})\Tr(a_1)}{N}+\frac{\Tr(a_2^{\dagger})\Tr(a_2)}{N}-1]^2+H_{\text{thermal}}
\end{equation}
with $J\sim 2\omega\ln{N}$. The ground space of this Hamiltonian is spanned by the following two states,

\begin{equation}\label{eqlls}
\begin{split}
    |\tilde{\uparrow}\rangle&=\frac{\Tr(a_1^{\dagger})}{\sqrt{N}}|0\rangle_{12}\\
    |\tilde{\downarrow}\rangle&=\frac{\Tr(a_2^{\dagger})}{\sqrt{N}}|0\rangle_{12} .
\end{split}
\end{equation}
where $|0\rangle_{12}$ is again the tensor product of the vacuum states of the two oscillators. These states span the code subspace of the model. 

We assume that $H_{\rm thermal}$ contains only single trace operators:
\begin{equation}\label{eqloth}
    H_{\rm thermal}=\sum_{\{n_k\}} \lambda_{\{n_k\}} b_{\{n_k\}} \frac{:\Tr(P\{a_1^{\dagger n_1}a_2^{\dagger n_2}a_1^{n_3}a_2^{n_4}\}):}{N^{\frac{\sum_{k=1}^4 n_k}{2}}}+h.c.  .
\end{equation}
where $P$ again denotes the sum over distinct arrangements of the operators inside the curly bracket, and all the operators are then normal ordered. In general, $\Tr(P\{a_1^{\dagger n_1}a_2^{\dagger n_2}a_1^{n_3}a_2^{n_4}\})$ is a sum over $\frac{(n_1+n_2+n_3+n_4)!}{n_1!n_2!n_3!n_4!}$ terms.  

The interaction term in the Hamiltonian Eq.~\eqref{eqHlocal} incurs an energy penalty for any logical errors that involve $\Tr(a_1^{\dagger})$ or $\Tr(a_2^{\dagger})$, so their probability to appear are suppressed due to the thermal factor.  For example, the probability of the error operator $\Tr(a_1^{\dagger})\Tr(a_2)$ is $\gamma(J)\gamma(-J)\sim e^{-\frac{J}{T}} $.
We set the energy scale $J$ to be 2$\omega \ln{N}$, which  guarantees that these uncorrectable errors are unlikely to occur when $N$ is large enough. All the other operators generated by the coupling in Eq.~\eqref{eqloth} are treated as possible errors that could act on the logical state. To show that these errors are approximately correctable, and to invoke Theorem~\ref{thm}, it suffices to check that aKL holds for single-trace errors, $E_{\{n_k\}}:=\frac{:\Tr(P\{a_1^{\dagger n_1}a_2^{\dagger n_2}a_1^{n_3}a_2^{n_4}\}):}{N^{\frac{\sum_{k=1}^4 n_k}{2}}}$,
\begin{equation}\label{eqn:localaKL}
\begin{split}
    \langle\tilde{i}|E_{\{n_k'\}}^{\dagger}E_{\{n_k\}}|\tilde{j}\rangle&=f_{\{n_k'\},\{n_k\}}\delta_{ij}+\frac{g_{\{n_k'\},\{n_k\}}^{ij}}{N^2}+O(\frac{1}{N^4})\\
    \langle\tilde{i}|E_{\{n_k\}}|\tilde{j}\rangle&=\frac{e^{ij}_{\{n_k\}}}{N}+O(\frac{1}{N^3}).
\end{split}
\end{equation}

Applying the argument we discussed in Section~\ref{case2}, one can again derive the aKL equation for general multi-trace operators using the above relations for single-trace operators. This guarantees that the aKL condition holds for all the error operators generated in this model. A proof of the aKL condition for the most general error is given in Appendix~\ref{aploaKL}. Here we discuss a few helpful examples to show that these errors indeed conform with the aKL condition.

Again, let us first check operators of the form $\frac{\Tr(a_{1,2}^{\dagger}a_{1,2})}{N}$. 
They induce a  phase error of size $O(\frac{1}{N})$,
\begin{equation}
\begin{split}
    &\langle\tilde\uparrow|\frac{\Tr(a_1^{\dagger}a_1)}{N}|\tilde{\uparrow}\rangle=\frac{1}{N}\\
    &\langle\tilde\downarrow|\frac{\Tr(a_1^{\dagger}a_1)}{N}|\tilde{\downarrow}\rangle=0,
\end{split}
\end{equation}
which is consistent with the aKL condition in Eq.~\eqref{eqn:localaKL}. Next, we check the aKL conditions for single-trace errors of the form $E_{(n,m)}=\frac{\Tr(a_1^{\dagger n}a_2^{\dagger m})}{N^{\frac{m+n}{2}}}$, with $n\geq 1,m\geq1$. For the bit flip errors, we can show that 
\begin{equation}\label{eqh2}
    \langle \tilde{\downarrow}|E_{(n'+1,m’-1)}^{\dagger}E_{(n,m)}|\tilde{\uparrow}\rangle =\delta_{nn'}\delta_{mm'}\frac{(n+1)m}{N^2}+O(\frac{1}{N^4}),
\end{equation}
and for the phase errors, we have
\begin{equation}\label{phase2}
    \begin{split}
        &\langle \tilde\uparrow|E_{(n',m')}^{\dagger}E_{(n,m)}|\tilde\uparrow\rangle =\delta_{nn'}\delta_{mm'}[1+\frac{n^2-m^2}{N^2}+O(\frac{1}{N^4})]\\
       & \langle \tilde\downarrow|E_{(n',m')}^{\dagger}E_{(n,m)}|\tilde\downarrow\rangle =\delta_{nn'}\delta_{mm'}[1+\frac{m^2-n^2}{N^2}+O(\frac{1}{N^4})].
    \end{split}
\end{equation}
When $m$ or $n$ equals to $0$, the equations are slightly different. For instance, when $m=0, n\geq 2$, the matrix elements are,
\begin{equation}
\begin{split}
    &\langle \tilde\uparrow|E_{(n',0)}^{\dagger}E_{(n,0)}|\tilde\uparrow\rangle =\delta_{nn'}[1+\frac{n(n-1)}{N^2}+O(\frac{1}{N^4})]\\
    &\langle \tilde\downarrow|E_{(n',0)}^{\dagger}E_{(n,0)}|\tilde\downarrow\rangle =\delta_{nn'}\\
    &\langle\tilde{\uparrow}| E_{(n',0)}^{\dagger}E_{(n,0)} |\tilde{\downarrow}\rangle=0.
\end{split}
\end{equation}

Although $H_{\rm thermal}$ only contains single-trace operators, the expansion in Eq.~\eqref{expansion} involves multi-trace operators acting on the state. To see the effect of these multi-trace errors, we evaluate an example:
\begin{equation}
\begin{split}
    &\langle \tilde\uparrow|E_{(p,0)}^{\dagger}E_{(q,0)}^{\dagger}E_{(m,0)}E_{(n,0)}|\tilde\uparrow\rangle =F(m,n,p,q)+\frac{G(m,n,p,q)}{N^2}+O(\frac{1}{N^4})]\\
    &\langle \tilde\downarrow|E_{(p,0)}^{\dagger}E_{(q,0)}^{\dagger}E_{(m,0)}E_{(n,0)}|\tilde\downarrow\rangle =F(m,n,p,q)+O(\frac{1}{N^4}).
\end{split}
\end{equation}
If the Knill-Laflamme conditions are satisfied, then the above expressions should yield matrix elements that only depend on the errors. We see that the presence of the $G$ function does allow one to distinguish the logical states, but the effect is suppressed by $1/N^2$. These correlation functions both have simple  diagrammatic representations similar to those shown in Figure~\ref{fig2} and Figure~\ref{fig3}, but the explicit form of these two functions are complicated, 
\begin{equation}\label{eqg}
\begin{split}
&F(m,n,p,q)=(\delta_m^p\delta_n^q+\delta_m^q\delta_n^p)+\delta_{m+n}^{p+q}\frac{f(p,q,m,n)}{N^2}\\
&G(m,n,p,q)=(\delta_m^p\delta_n^q+\delta_m^q\delta_n^p)[n(n-1)+m(m-1)]+(\delta_{m\pm1}^p\delta_{n\mp 1}^q+\delta_{m\pm 1}^q\delta_{n\mp 1}^p)\sqrt{pqmn}.
\end{split}
\end{equation}
with some function $f$ whose exact form is not essential for our argument.

Again all the errors discussed above are approximately correctable errors, because they can only corrupt the logical information up to  $O(\frac{1}{N^2})$. The uncorrectable errors or logical operators are those that contain $\Tr(a_1^{\dagger})$ and $\Tr(a_2^{\dagger})$. However, recall that we have tuned the coupling $J$ to be proportional to $2\omega\ln{N}$ so that these uncorrectable errors have a similar $1/N$ suppressed effect comparable in size to the approximately correctable errors.

By construction, the bath model has uniform coupling as required by condition 2 of Theorem~\ref{thm}. In addition, the number of gauge invariant error operators with energy $n\omega$ that are induced by Eq.~\eqref{eqloth} grows asymptotically as $2^n$. In particular, the number of effective error operators that act non-trivially on the logical state, i.e., ones contributing to the information leak in the Lindblad equation, must have $n_2,n_4\leq 1$. Therefore, the total number of effective errors with energy $\omega n$ is bounded by an exponential function, as required by condition 1. 

Because the aKL condition holds and the bath model only generates errors which are protected by the code, we again invoke the Theorem~\ref{thm}. The mutual information between the encoded qubit $S$ and a reference $R$ is
\begin{equation}
I(S:R)(t)=2\ln{2}-K_{\rm local}(\frac{t}{N^2}).
\end{equation}
The expression $K_{\rm local}(t)$ in this model is slightly different from $K(t)$, since the $O(\frac{1}{N^2})$ term in the aKL condition depends quadratically on $n$ (see Eq.~\eqref{eqh2}).  When $\lambda t \ll N^2$, $K_{\rm local}(\frac{t}{N^2})\sim \frac{\sum_n n^2\lambda_n t}{N^2}$.  At late time $\lambda t\sim N^2$, $K_{\rm local}(t)\sim (\sum_n n^2 \lambda_n t)^2$.  

 More importantly, note that the rate of information loss is given by
\begin{equation}
\begin{split}
    \sum_n n^q\lambda_n=&\sum_n n^qd_ne^{-n \beta \omega}\\
    \sim& \sum_n n^q2^n e^{-n\beta \omega},
\end{split}
\label{eqn:hagedorn}
\end{equation}
where $q=2$, and $d_n\sim 2^n$ is the number of single trace states at the $n$th energy level. A crucial distinction from the previous model is the Hagedorn-like transition with a critical temperature $T_c=\frac{\omega}{\ln{2}}$ \cite{Aharony:2003sx}. In the low temperature phase where $T<T_c$, the rate of information loss \eqref{eqn:hagedorn} is finite and $N$-independent. Therefore the large $N$ gauged matrix model has a long memory time $\tm \sim N^2$.
However, for $T>T_c$, \eqref{eqn:hagedorn} exhibits a Hagedorn-like behaviour\footnote{Note that for finite $N$, the Hagedorn behaviour does not continue to arbitrary energies, but only to order $N^2$, the total number of gauge invariant degrees of freedom. The counting of gauge invariant operators can be found in Section~\ref{MIdiag}. }. As a result, $K_{\rm local}(t)$ now also scales with $N$. This $N$ dependence cancels out the large $N$ suppression, and the memory time no longer scales with the system size. This behaviour is similar to the holographic code \cite{thresholdadscft}, where the temperature needs to be lower than a critical value  to maintain the robustness of the encoded information.

Since the logical states are within the ground space of this model, there are no operators that can lower the energy. The zero energy error operators do not contribute to information loss at zero temperature limit, due to the vanishing thermal spectral function $\lim_{\beta\rightarrow \infty} \gamma(\nu=0)$. Therefore, when $\beta\rightarrow \infty$, we expect the information decay rate to scale as $\frac{1}{N^2}e^{-\beta\omega}$ according to Eq.~\eqref{eqn:hagedorn}.



 \section{Relaxing the gauge constraint }\label{case3}
 
 We now relax the gauge constraint to a global symmetry, which we then impose energetically. The purpose of this is two-fold --- (i) we want to understand how the stability of quantum memory is tied to the gauge symmetry specifically, and (ii) there are very few models in nature that are known to have gauge symmetries. For example, Quantum ElectroDynamics (QED) has a $U(1)$ symmetry and  Quantum ChromoDynamics (QCD) has an $SU(3)$ symmetry. However, there is no known $SU(N)$ gauged model that can be realized in the laboratory, especially for arbitrarily large $N$. Therefore, in this section, we consider a single oscillator model with only a global $SU(N)$ symmetry.  The system of interest has Hamiltonian
 \begin{equation}\label{eqcasi}
 \begin{split}
     H_{\rm Gauss}=\frac{\mathcal{J}}{N} \sum_{ij}G^i_{j}G^{j}_i\\
     G^i_j=\sum_k a^{\dagger i}_k a^{ k}_j-a^{\dagger k}_ja^i_k,
\end{split}
 \end{equation}
 where $G^i_j|\psi\rangle =0$ is the Gauss law constraint associated with the gauge symmetry. The energy penalty term plays the role of a Casimir operator in the group $SU(N)$, with eigenvalues proportional to $\mathcal{J}$ and being the same for all states in an irreducible representation. 
 This term in the Hamiltonian forces the low energy states to be singlets of the symmetry group. 
 
In the absence of gauge symmetry, we allow the  modes $a^{ i}_j$'s to couple  independently with the thermal  bath operator. Let us consider the following thermal coupling Hamiltonian,
 
 \begin{equation}
 \begin{split}
     H_{\rm thermal}=&\lambda_1\sum_{ij}b^{ i}_ja^{ j (\dagger)}_i+\lambda_2\sum_{ijkl}b^{ ik}_{jl} a^{  j (\dagger)}_{i}a^{ l (\dagger)}_k+\lambda_3\sum_{ijklpq}b^{ikp}_{jlq}a^{j (\dagger)}_ia^{l (\dagger)}_ka^{q (\dagger)}_p+\cdots\\
     &+h.c.  .
\end{split}
\label{eqn:nongaugecoupling}
\end{equation}
 
One can show that the Lindblad Master equation takes the form of 
 
 \begin{equation}\label{eqLind}
     \dot{\rho}(t)=\lambda_1\frac{\Tr(a^{\dagger})}{\sqrt{N}}\rho\frac{\Tr(a)}{\sqrt{N}}+\lambda_1e^{-\frac{\mathcal{J}}{T}}\sum_{ij}\tilde{a}^{\dagger i}_j\rho \tilde{a}^j_i+\cdots ,
 \end{equation}
 where $\tilde{a}^i_j$ represents the component that transforms in the adjoint representation of $SU(N)$, such that
 \begin{equation}
     \tilde{a}^{\dagger i}_j=a^{\dagger i}_j-\frac{\Tr(a^{\dagger})}{N}\delta^i_j.
 \end{equation}
We show in Appendix~\ref{apLind} that it is a consequence of the $H_{\rm Gauss}$ term in the Hamiltonian.
 
These non-singlet errors acting on the ``gauge invariant'' logical states, \textit{i.e.}, logical states satisfying the Gauss law constraint, only corrupt quantum information at $O(\frac{1}{N^2})$, which we show in Appendix~\ref{globaleg}. The summation in Eq.~\eqref{eqLind} is over $N^2$ number of such errors and therefore the total effect of these non-singlet errors  is proportional to $N^2e^{-\frac{\mathcal{J}}{T}}$.  Hence the coupling strength $\mathcal{J}$ has to be at least $\sim 2T\log{N}$ in order to suppress their effects to the same order as those from the the singlet errors. Note that this model effectively contains one such matrix oscillator and does not exhibit a Hagedorn-like transition. Nevertheless, it is clear that stability is only ensured when $T\lesssim \mathcal{J}$, hence can also depend on the system size\footnote{This should be contrasted with the models with exact gauge symmetry, where the temperature that sets apart a stable and an unstable memory only depends on the natural frequency of the system. Note that in the local model with gauge symmetry, there is an additional scale introduced by $J$ which has to satisfy $J\gtrsim \omega \log N$ to to ensure stability. However, as long as we have tuned $J>T_c$, then $T_c$ will be the relevant energy scale for the low energy stability.}. In the following, we will assume their suppression with a strong enough coupling. As for the singlet error operators, they are defined as,
\begin{equation}\label{eqglerror}
    E_{(\{n_k,m_k)\}}:=\prod_{k} \frac{\Tr(P\{a^{\dagger n_k}a^{m_k}\})}{N^{\frac{\sum_k n_k+m_k}{2}}}.
\end{equation}
 
The logical states we choose in this model are (divided by proper normalization constant),
 \begin{equation}
 \begin{split}
     &|\tilde{1}\rangle =\frac{\Tr(a^{\dagger L})^2}{\mathcal{N}(L)}|0\rangle  \\
     &|\tilde{2}\rangle =\frac{\Tr(a^{\dagger L+1})\Tr(a^{\dagger L-1})}{\mathcal{N}(L)}|0\rangle .
\end{split}
 \end{equation}
We also define the orthogonormal basis,
\begin{equation}\label{eqglbasis}
    \begin{split}
        & |\tilde{\uparrow}\rangle=\frac{|\tilde{1}\rangle+|\tilde{2}\rangle}{2(1+\langle\tilde{1}|\tilde{2}\rangle)}\\
        & |\tilde{\downarrow}\rangle=\frac{|\tilde{1}\rangle-|\tilde{2}\rangle}{2(1-\langle\tilde{1}|\tilde{2}\rangle)}.
    \end{split}
\end{equation}

In order to recast the code subspace as a lower energy subspace, let us consider a shifted version of the system Hamiltonian and write the full Hamiltonian as
 \begin{equation}
     H=\frac{\mathcal{J}}{N}\sum_{ij}G^i_jG^j_i+\omega(\Tr(a^{\dagger}a)-L)^2+H_{\rm thermal},
 \end{equation}
where $L>0$ is some arbitrary integer. 


One can check that the aKL condition,
 \begin{equation}
 \begin{split}
    \langle\tilde{i}|E_{\{(n_k',m_k')\}}^{\dagger}E_{\{(n_k,m_k)\}}|\tilde{j}\rangle&=f_{\{(n_k',m_k')\},\{(n_k,m_k)\}}\delta_{ij}+\frac{g_{\{(n_k',m_k')\},\{(n_k,m_k)\}}^{ij}}{N^2}+O(\frac{1}{N^4})\\
    \langle\tilde{i}|E_{\{(n_k,m_k)\}}|\tilde{j}\rangle&=\frac{e^{ij}_{\{(n_k,m_k)\}}}{N}+O(\frac{1}{N^3}),
    \end{split}
\end{equation}
holds for all relevant errors induced by \eqref{eqglerror} as long as they do not contain $\Tr(a^{\dagger L})$ or $\Tr(a^{\dagger L\pm 1})$ as a factor. 
For example, consider the class of error operators   $E_n=\frac{\Tr(a^{\dagger n})}{\sqrt{n}N^{\frac{n}{2}}}$. The relevant matrix elements are 

\begin{equation}
\begin{split}
  & \langle \tilde{\uparrow}|E_m^{\dagger}E_n|\tilde{\downarrow}\rangle =\delta_{mn}[\frac{2n}{N^2}+O(\frac{1}{N^4})]\\
   & \langle \tilde{\uparrow}|E_m^{\dagger}E_n|\tilde{\uparrow}\rangle-\langle \tilde{\downarrow}|E_m^{\dagger}E_n|\tilde{\downarrow}\rangle= O(\frac{1}{N^4}).
\end{split}
\end{equation} 

 Note that although the Hamiltonian and logical states depend on $L$, the error induced by a low energy operator has no such dependence (the $O(\frac{1}{N^2})$ term does not depend on $L$); ergo, the aKL conditon in the standard form Eq.~\eqref{eqknill} holds for such singlet errors (\ref{eqglerror}). We present the details of the proof in Appendix~\ref{globaleg}).

Similar to the discussion in Section~\ref{case2}, the number of singlet errors with the energy $n^2\omega$ scales as the number of integer partitions $p(n)$. This is clearly slower than exponential growth. Therefore, if the thermal coupling in (\ref{eqn:nongaugecoupling}) does not contain $\Tr(a^{\dagger L})$ and $\Tr(a^{\dagger L\pm1})$ as factors, and $\lambda \sim O(1)$, then Theorem~\ref{thm} applies and we again attain a memory time quadratic in $N$. Since the logical states are within the ground space, there are no spontaneous decay of quantum information. So the memory time is proportional to $N^2e^{\frac{\omega}{T}}$ at zero temperature limit. 

However, the low energy errors like $\Tr(a^{\dagger L})\Tr(a^L)$ do appear, which means that the aKL condition does not hold for these operators. Nevertheless, it is possible to retain this long memory time in this model as long as we also suppress the singlet errors containing such terms. One way to achieve this is to assume that the coupling constants are exponentially suppressed with respect to the number of modes involved, i.e., $\lambda_k\sim \lambda^k$ for some $\lambda<1$. 
 This gaps out the dangerous error for large $L\sim \log{N}$ by suppressing the error probability to order $\frac{1}{N^2}$. In the meantime, it is also a natural requirement in this model because the thermal bath mode couples to each oscillator individually.

One potential drawback in this model (and also the fully gauged model in Section~\ref{case1}) is that the coupling constants $\mathcal{J}, J\propto \ln N$ depend on the system size. However,  this is not as unreasonable as it looks because the interaction energy per harmonic oscillator mode is small\footnote{The energy for thermal state at temperature $T\sim \omega\ll \mathcal{J}$ is $O(1)$ times $\mathcal{J}$. See Appendix~\ref{apcasimir} for more detail. } and is proportional to $\frac{\ln{N}}{N^2}\omega$. For reference, let us contrast this $N$-dependence with that of the toric code model with system size equal to $N$. Recall that the toric code is a $Z_2$ lattice gauge theory and the Hamiltonian also involves terms that are proportional to the Gauss's law operators, with coupling constant $\Delta$. When it is coupled to a bath with fixed temperature $T$, the memory time is proportional to $\frac{1}{N}e^{\frac{\Delta}{T}}$. It becomes a viable quantum memory if one can tune the coupling $\Delta$ to $T\ln{N}$, which corresponds to an energy density much larger than that required in our matrix model.

 \section{Large N spin model}\label{spin}
 

In the previous sections, we saw that exact $SU(N)$ symmetries in oscillator models can lead to robust quantum memories, both when the symmetry is fully gauged and when the singlet restriction is enforced energetically. It is natural to ask whether such exact symmetries and infinite dimensional Hilbert spaces are truly necessary for robustness. In other words, can we realize an analogous phenomenon with finite dimensional systems, e.g. involving qubits, that only have an approximate $SU(N)$ symmetry? In this section, we propose one such model and show that the requisite large $N$ factorization properties are mostly preserved up to small corrections in the low energy sector. As a result, the general conclusions of Theorem \ref{thm} still apply to this system in a way that is similar to Section 4.
 
To motivate the model, let us first review how one can approximately embed the low-energy states of an oscillator into a spin Hilbert space. Consider a spin-$j$ representation of $SU(2)$. When acting on states with $S^z$ eigenvalue close to maximal, $m \lesssim j$, the spin commutator becomes
\begin{equation}
    [S^x,S^y] = i \hbar S^z \approx i \hbar^2 j + \cdots.
\end{equation}
This suggest that $S^x$ and $S^y$ can be approximately mapped to a canonical pair. Setting
\begin{equation}
    Q = \frac{\ell}{\hbar \sqrt{j}} S^x, \,\,\,\, P = \frac{1}{\ell \sqrt{j}} S^y 
\end{equation}
for some as yet undetermined length $\ell$, we see that 
\begin{equation}
    [Q,P]\approx i \hbar +\cdots.
\end{equation}

Now considering the harmonic oscillator Hamiltonian
\begin{equation}
    H = \frac{1}{2} (P^2 + \omega^2 Q^2) = \frac{1}{2} \left(\frac{1}{\ell^2 j} (S^y)^2  + \frac{\omega^2 \ell^2}{\hbar^2 j} (S^x)^2 \right),
\end{equation}
we may set $\ell^2 = \hbar/\omega$ to obtain
\begin{equation}
    H = \frac{\omega}{2 \hbar j} \left((S^x)^2 + (S^y)^2 \right) = \frac{\omega}{2 \hbar j} (S^2 - (S^z)^2).
\end{equation}
Using the standard spectrum of $S^2$ and $S^z$ with $m=j-\Delta$, the spectrum of $H$ is
\begin{equation}
    \frac{\hbar \omega}{2j}\left( j(j+1) - (j-\Delta)^2 \right) = \hbar \omega \left( \Delta + \frac{1}{2} + O(\Delta^2/j)\right).
\end{equation}

From this discussion, we expect that the physics of our oscillator models may be suitably extended by replacing the oscillators with qudits of dimension $d$ ($d=2j+1$ in the above mapping). At large $d$ we have a nearly identical low-energy spectrum, but the models make sense for any $d$ and we will show that even $d=2$ ($j=1/2$) suffices.

Specifically, we can construct a system of $N^2$ spins, which can be thought of as qudits with local dimension $d$. It is tempting to write each spin operator in the adjoint representation of $SU(N)$, meaning each spin operator carries indices $i,j$ running from $1$ to $N$ and transforms as
\begin{equation}\label{eqspintransform}
    S^{i\pm}_j\rightarrow \sum_{kl}U^i_kS^{k\pm}_lU^{\dagger l}_j.
\end{equation}
Each spin also satisfies the following $SU(2)$ algebra,
\begin{equation}
\begin{split}
   &[ S_{i}^{j+},S_j^{i -}]=2S_{ i}^{j z}\\
   &[S_{ i}^{j z},S_i^{j+}]=S_i^{j+}\\
   &[S_{i}^{j z},S_j^{i-}]=-S_j^{i-},
\end{split}
\end{equation}
where $i,j$ again label the different spins or qudits. However, the $SU(N)$ symmetry does not act consistently on the spin algebra due to the first line above: we cannot interpret $S_i^{jz}$ as being in the adjoint of $SU(N)$ while also obeying the commutation relation. Moreover, even if we grant that $S_i^{jz}$ is not in the adjoint representation, it is impossible to construct $SU(N)$ generators $G_i^j$ with the desired commutation relations with $S_i^{j \pm}$, i.e. there is no $G_i^j$ obeying\footnote{The inconsistency of $SU(N)$ and the $SU(2)$ algebra can be seen from the equation $\sum_k[[S^{i+}_k,S^{k-}_j],S^{l+}_h]=\delta^i_j\delta^j_hS^{l+}_h$, where the right hand side transforms differently from the left hand side.  }

\begin{equation}
    [G_i^{j},S_k^{l\pm}] = \delta_i^l S_k^{j\pm}-\delta_k^j S_i^{l\pm}.
\end{equation}

However, it is still possible to realize an approximate symmetry in terms of the pseudo-generators, \begin{equation}\label{eqapproxalgebra}
\begin{split}
    \tilde{G}_i^j=\sum_k S^{j+}_kS^{k-}_i-S^{k+}_iS^{i-}_k\\
[\tilde{G}_i^j,S_k^{l\pm}]=(\delta_i^l S_k^{j\pm}-\delta_k^j S_i^{l\pm})S^{lz}_k.
\end{split}
\end{equation}
Indeed, if $S^z$ could be approximated as a constant operator, then we would have the desired commutators. This approximate algebra thus connects back with the oscillator models of the previous sections where the symmetry becomes exact. In what follows, we work with the pseudo-generators $\tilde{G}_i^j$ and set $d=2$.

Similar to our recipe in the previous sections, one can construct the analogs of the gauge invariant states by acting multi-trace operators on the tensor product of spin down states 
\begin{equation}
\begin{split}
   &\Tr(S^{+ n_1})\Tr(S^{+ n_2})\cdots \Tr(S^{+ n_k}) |0\rangle \\ &|0\rangle :=|\downarrow\downarrow\cdots\downarrow\rangle .
\end{split}
\end{equation}
For convenience, we still refer to these analogues of the gauge invariant operators (states) singlet operators (states).
Indeed, one can see that the desired symmetry is only approximate because the algebra of such operators is not closed. For example,
\begin{equation}
    \Tr(S^{- 2})\Tr(S^{+ 2})\Tr(S^{+ 2})|0\rangle =(4N^2-4N-8)\Tr(S^{+ 2})|0\rangle +32\sum_iS^{i+}_iS^{i+}_i|0\rangle.
\end{equation}
Generally, the product of such singlet terms can produce non-singlet terms that are only sub-leading in $N$ \footnote{If we further restrict ourselves to only consider spin-$\frac{1}{2}$,i.e., $d=2$, then the non-singlet terms also vanishes. This can be a convenient simplification when the local systems are exactly qubits, however, our analysis of robustness will also hold for systems with $d>2$.}.  

Now let us consider a system Hamiltonian $H_S=hH_0+\mathcal{J} H_G$, where
\begin{equation}
    H_0=(\sum_{ij}S^{iz}_j-S_{\rm tot})^2,
\end{equation}
and $-\frac{N^2}{2}\leq S_{\rm tot}\leq \frac{N^2}{2}$ is some $N$-dependent offset. $H_G$ is an energy penalty term similar to the Gauss law constraint; it corresponds to the Casimir operator that enforces the approximate $SU(N)$ symmetry for low energy states\footnote{The Casimir operator $C=\sum_{ij}G^i_jG^j_i$ for an exact $SU(N)$ symmetry has a spectrum of the form $NwJ$, where $w$ is the highest weight of given representation. Analogously, the Hamiltonian $H_G$ also has an energy spectrum of the form of $wJ$ up to $\frac{1}{N}$ corrections. }, 
\begin{equation}\label{eqHG}
     H_G= \frac{1}{N}\tilde{G}^2:= \frac{1}{N}\sum_{ij}\tilde{G}_i^j\tilde{G}_j^i.
\end{equation}
When $\mathcal{J}$ is large, it creates a large gap in the  spectrum. We define the low energy sector as the states with eigenvalues suppressed by $\frac{1}{N}$. The energy eigenstates of $H_S$ are generally superpositions of the singlet and non-singlet states, but it can be shown that the low energy sector is isomorphic to the singlet subspace. For example, the following are four states created by product of  three $S^+$'s, which are simultaneous eigenstates of both $H_0$ and $H_G$,
\begin{equation}
\begin{split}
    &|\psi_1\rangle =\Tr(S^{+3})|0\rangle \\
    &|\psi_2\rangle =[\Tr(S^+)^3-3\Tr(S^{+2})\Tr(S^{+})]|0\rangle \\
    &|\psi_3\rangle =[\Tr(S^{+2})\Tr(S^{+})+(1+\frac{2}{N})\Tr(S^+)^3-\frac{19}{2N}\sum_i(S^{+2})^i_iS^{+i}_i]|0\rangle \\
    &|\psi_4\rangle =[\sum_i(S^{+2})^i_iS^{+i}_i-\frac{1}{N}\Tr(S^{+2})\Tr(S^+)]|0\rangle .
\end{split}
\end{equation}

The first three states are in the low energy sector, since they are eigenstates of $H_G$ with  eigenvalues $E_1^G=E_2^G=0$ and $E_3^G=\frac{7\mathcal{J}}{2N}$. The fourth state has eigenvalue $E_4^G=\mathcal{J}$, and hence a high energy excitation. 

There are two potential causes for this model to deviate from the one in Section~\ref{case3} --- effects from the symmetry being approximate and finite dimensionality of each spin.  Formally, we should still expect the low energy states, \textit{i.e.}, ones with energy $\ll \mathcal J$, to have a one-to-one correspondence with the multi-trace operators formed by $S^+$'s. More precisely, their low energy spectra should coincide up to corrections that are $1/N$ suppressed and the multiplicity, when we ignore such corrections, should be identical.

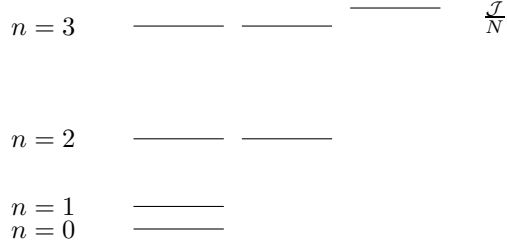
\begin{figure}[H]
    \centering
   \begin{tikzpicture}[scale=0.3]
   \draw (-4,0) node {$n=0$};
   \draw (-4,1) node {$n=1$};
   \draw (-4,4) node {$n=2$};
   \draw (-4,9) node {$n=3$};
   \draw (16,9.4) node {$\frac{\mathcal{J}}{N}$};
   \draw (0,0)--(4,0);
    \draw (0,1)--(4,1);
    \draw (0,4)--(4,4);
    \draw (4.8,4)--(8.8,4);
    \draw (0,9)--(4,9);
    \draw (4.8,9)--(8.8,9);
    \draw (9.6,9.8)--(13.6,9.8);
   \end{tikzpicture}
    \caption{Specturm of the low energy states. The ground state has energy zero. Excited states has energy proportional to  $n^2h$.  Note that due to the non-exactness of symmetry, there are small energy splitting proportional to $\frac{\mathcal{J}}{N}$ . For sufficiently large $N$, this small gap can be neglected and  the degeneracy at each level is the same as in harmonic oscillator model. Non-singlet excitations have energy $O(\mathcal{J})$, which we do not count as low energy.}
    \label{fig5}
\end{figure}

We see this by noting that their energy only depend on number of $S^+$ involved, with small corrections proportional to $\frac{\mathcal{J}}{N}$ (Figure~\ref{fig5}). Therefore, up to these small corrections introduced by the deviations from the exact symmetry, the operator counting at each energy level is the same as that in Section~\ref{case3}. These corrections can be connected to the break down of the exact symmetry. 

As for the correction from finite dimensionality,  thanks to the ``Gauss law'' constraint $H_G$ and the form of $H_0$, these corrections would only significantly modify this counting when the total number of $S^+$ involved is of $O(N)$. Therefore, in the low energy regime where we operate, both effects are suppressed by having a large enough $N$. 

Similar to the harmonic oscillator model,  we consider the full Hamiltonian, 
\begin{equation}
    H=hH_0+\mathcal{J}H_G+H_{\rm thermal}
\end{equation}
with $H_{\rm thermal}$ being

 \begin{equation}
 \begin{split}
     H_{\rm thermal}=&\lambda_1\sum_{ij}b^{ i}_jS^{ j \pm}_i+\lambda_2\sum_{ijkl}b^{ ik}_{jl} S^{  j \pm}_{i}S^{ l \pm}_k+\lambda_3\sum_{ijklpq}b^{ikp}_{jlq}S^{j \pm}_iS^{l \pm}_kS^{q \pm}_p+\cdots\\
     &+h.c. .
\end{split}
\end{equation}

The lowest energy states must have total spin equal to $S_{\rm tot}$. Let us define $L=\frac{S_{\rm tot}+N^2/2}{2}$ and pick a two dimensional subspace as the code subspace. With proper normalization, 
\begin{equation}
\begin{split}
    |\tilde{1}\rangle &=\frac{\Tr(S^{+ L})^2}{\mathcal{N}}|0\rangle \\
    |\tilde{2}\rangle &=\frac{\Tr(S^{+ L+1})\Tr(S^{+ L-1})}{\mathcal{N}}|0\rangle .\\
\end{split}
\end{equation}

Again, we define the orthogonormal basis in this logical subspace as, 
\begin{equation}
    \begin{split}
        & |\tilde{\uparrow}\rangle=\frac{|\tilde{1}\rangle+|\tilde{2}\rangle}{2(1+\langle\tilde{1}|\tilde{2}\rangle)}\\
        & |\tilde{\downarrow}\rangle=\frac{|\tilde{1}\rangle-|\tilde{2}\rangle}{2(1-\langle\tilde{1}|\tilde{2}\rangle)}.
    \end{split}
\end{equation}
with the overlap $\langle\tilde{1}|\tilde{2}\rangle=\frac{f(L)L\sqrt{L^2-1}}{\sqrt{2}N^2}$, where $f(L)$ is a complicated function defined in Appendix~\ref{apspin}.

Let us couple this system to a thermal bath at temperature $T$. First note that the non-singlet excitations are again suppressed by a factor of $e^{-\frac{\mathcal{J}}{T}}$ because of the energy penalty term $H_G$, while the singlet interaction has no such suppression. Then let us reorganize the system-bath interactions by the approximate symmetry such that
\begin{equation}
\begin{split}
    H_{\rm thermal}&=\lambda_1\sum_{ij}b_j^i S^{j\pm}_i+\lambda_2\sum_{ijkl}b^{ik}_{jl}S^{j\pm}_iS^{l\pm}_k+ \cdots+h.c.\\
    &=\lambda_1\sum_{ij}b_j^i(S_i^{j+}-\frac{\Tr(S^+)}{N}\delta_i^j)+\lambda_1\frac{\Tr(b)}{\sqrt{N}}\frac{\Tr(S^+)}{\sqrt{N}}+\cdots .
\end{split}
\end{equation}
 Similar to the discussion in last section, this interaction induces an order $N^2$ number of non-singlet errors. Therefore, the total error contribution from non-singlet excitations is $\sim N^2e^{-\frac{\mathcal J}{T}}$ times that of singlet excitations. To suppress both types of errors to the same order, we require the coupling $\mathcal{J}$ to be greater than $2h\log{N}$ . 


More precisely, the singlet error operators are defined as 

\begin{equation}\label{eqspinerror}
    E_{(\{n_k,m_k)\}}:=\prod_{k} \frac{\Tr(P\{S^{+ n_k}S^{- m_k}\})}{N^{\frac{\sum_k n_k+m_k}{2}}}.
\end{equation}

Again, we show that the aKL condition

 \begin{equation}
 \begin{split}
    \langle\tilde{i}|E_{\{(n_k',m_k')\}}^{\dagger}E_{\{(n_k,m_k)\}}|\tilde{j}\rangle&=f_{\{(n_k',m_k')\},\{(n_k,m_k)\}}\delta_{ij}+\frac{g_{\{(n_k',m_k')\},\{(n_k,m_k)\}}^{ij}}{N^2}+O(\frac{1}{N^4})\\
    \langle\tilde{i}|E_{\{(n_k,m_k)\}}|\tilde{j}\rangle&=\frac{e^{ij}_{\{(n_k,m_k)\}}}{N}+O(\frac{1}{N^3})
\end{split}
\end{equation}
is satisfied in Appendix~\ref{apspin}. For example, one can see that the operator $\frac{\Tr(S^{+}S^{-})}{N}=\frac{\sum_{ij}S^{i z}_j}{N}$ acts as the logical identity,  since the logical states have the same total spin in $z$ direction. For another example, the aKL conditions for errors $E_n=\frac{\Tr(S^{+ n})}{\sqrt{n}N^{\frac{n}{2}}}$ are

\begin{equation}\label{KLglobal}
\begin{split}
     &\langle \tilde\uparrow|E_m^{\dagger}E_n|\tilde\downarrow\rangle=\delta_{mn}[\frac{2n}{N^2}+O(\frac{1}{N^4})]\\
    &\langle \tilde\uparrow|E_m^{\dagger}E_n|\tilde\uparrow\rangle-\langle \tilde\downarrow|E_m^{\dagger}E_n|\tilde\downarrow\rangle=O(\frac{1}{N^4}).
\end{split}
\end{equation}

Note that in these examples, and more generally in Appendix~\ref{apspin}, deviations from the KL condition only depend on $n$ but not on $L$. Therefore, the aKL condition remains valid even for very large $L$. Similar to the last section, we can pick $L\gtrsim \log{N}$ to make sure uncorrectable errors are suppressed for large $N$. For instance, one may consider choosing $L=N^2/4$ such that $S_{\rm tot}=0$. 

Since this spin model has a structure very similar to the ungauged harmonic oscillator model, we can follow the same  argument in Section~\ref{case3}.  In the low energy sector with $\epsilon \ll \mathcal J\sim \log N$, (approximate) singlet errors are the dominant error modes. The number of such error operators grows sub-exponentially with energy because these operators still have the  multi-trace structure whose counting are the same as the one in Section 4. There are small corrections because of the approximate symmetry, however they are proportional to $\frac{\mathcal{J}}{N}$, and therefore for large $N$, they do not affect the overall structure of the spectrum for the purpose of our analysis (Figure~\ref{fig5}).  Similarly, we also need the thermal coupling  strength to decay exponentially with number of  spins to suppress the singlet error operators such as $\Tr(S^{-L})\Tr(S^{+ L})$. As argued in the previous section, this is a natural assumption since the bath mode couples to each spin individually. Since the rest of the argument is identical to those in Section~\ref{case3}, we can then invoke Theorem~\ref{thm} to conclude that the memory time $t_m$ scales quadratically with $N$. There is no operator that can lower the logical states' energy, so the zero temperature limit of memory time is still proportional to $N^2e^{\frac{\omega}{T}}$. We remark that this model also has the same drawback as the ungauged harmonic oscillator model, that is, the interactions remain non-local and we need the coupling strength $\mathcal{J}$ to scale as at least $\log{N}$.

\section{Discussion and Conclusion}
\label{sec:discussion}


Motivated by connections between error correction and phenomena in gauge theory and quantum gravity, we have taken a first step in understanding how $SU(N)$ gauge symmetries at large $N$ can be used to construct stable quantum memories. In particular, we construct a  gauged matrix quantum mechanics toy model with non-local logical operations that is a self-correcting quantum memory. The memory time of this model is polynomial in $N$ when coupled to a thermal bath at non-zero temperature. The presence of gauge symmetry induces approximate quantum error correcting codes that become exact in the large $N$ limit, thanks to the large $N$ factorization property. Furthermore, gauge invariance limits the number of error patterns that can be excited up to a certain energy scale. The combination of these factors suppresses the proliferation of memory errors under thermal noise. A similar model with local logical operation is also possible, but as a trade off, we impose an additional energy penalty with a coupling of strength $\sim \log N$. The memory time is again polynomial in $N$, but only for temperatures below a critical value.

As large $N$ gauge symmetries are not known to occur in nature, we relax the gauge symmetries and impose only large $N$ global symmetries energetically on a single matrix oscillator. We also construct a non-local finite dimensional spin model which further relaxes the symmetry constraint where we only enforce approximate large $N$ symmetries energetically. Interestingly, these constructions remain robust in both scenarios under the assumed system-bath coupling as long as the symmetry constraints can be energetically enforced with a penalty that scales as $O(\log N)$.

Stepping back from these particular examples, our work provides one set of guidelines for building large $N$ passive quantum memories in which the encoded information remains robust when coupled to a thermal bath. Sufficient conditions on the quantum system and its coupling to the bath are captured in Theorem~\ref{thm}. We hope to extend our current work in a number of future directions. Let us roughly categorize these potential generalizations in the direction of optimality, dimensionality, symmetries, variety, and practicality.

Although we can construct models that have $\text{poly}(N)$ memory time, they also often come with other features that are practically difficult, such as non-local interactions and $\log N$ coupling strengths. However, we by no means claim that these are the optimal constructions. Indeed, it is possible that a more careful design can produce better self-correcting memories. For example, an interesting theoretical problem is whether there exists a gauged $SU(N)$ oscillator model with local coupling for which the quantum memory is stable against thermal noise. 

It is also natural to generalize these findings to systems with gauge symmetries in higher dimensions by considering geometrically coupled oscillators, lattice gauge theories, and concatenations with other quantum error correcting codes. We will take a first step towards tackling a subset of these problems in an upcoming work \cite{fieldpaper}. Although we enforce global symmetries energetically in our work, one can similarly ask whether there are advantages in considering systems with emergent gauge symmetries at low energies, which is better understood in higher dimensional systems. Relatedly, one can examine what aspects of the $SU(N)$ symmetry are actually essential. For instance, does a similar result hold for discrete, instead of continuous, non-abelian symmetries?

Drawing intuition from gauge theories, we can also examine how to generalize such models by coupling the ``pure gauge'' degrees of freedom to fundamental degrees of freedom like matter fields and defects. In particular, how does the configuration of these other degrees of freedom impact our conclusions? One may, for instance, introduce a variety of fields that may interact under these gauge couplings. Extensions in this direction draw heavy parallels with traditional theoretical particle physics, and can open up new connections with quantum information in this area. 

Lastly, as a practical aspect of any quantum memory, information needs to be encoded and extracted relatively easily. This is yet unexplored for both the infinite dimensional oscillator and the finite dimensional spin system. As a standalone quantum code, the gauged matrix oscillator is a continuous variable system that resembles bosonic codes~\cite{gkp} in many ways. However, we have barely begun to analyze its code properties. It is desirable to understand its similarities and differences with existing bosonic codes, its robustness against different kinds of errors, and how active error correction like syndrome extraction can be performed.

\section*{Acknowledgement}
We thank Alexey Milehkin, John Preskill for helpful comments and discussions. C.C. acknowledges the support by the U.S. Department of Defense and NIST through the Hartree Postdoctoral Fellowship at QuICS, the Air Force Office of Scientific Research (FA9550-19-1-0360), and the National Science Foundation (PHY-1733907). G.C. acknowledges support from the U.S. Department of Energy, Office of Science, Office of Advanced Scientific Computing Research, Accelerated Research for Quantum Computing program ``FAR-QC''. B.G.S. acknowledges support from the AFOSR under grant number FA9550-19-1-0360. 

\bibliographystyle{unsrt} 
\bibliography{ref} 

\appendix
\section{Proof of Theorem \ref{thm}}\label{apentropy}

In this section, we will prove that the mutual information decays polynomially in $t$ when coupled to a thermal bath of suitably low temperature, as long as the conditions in Theorem~\ref{thm} are satisfied. We first set up some useful notations for later use.

Let $\cal{E}$ be the set of operators  that generate the errors. For all operators $E_a\in \mathcal{E}$ satisfying the aKL equation, we can always  diagonalize the matrix $f_{ab}$ by a basis transformation inside $\mathcal{E}$, such that $\langle\tilde{i}| E_b^{\dagger}E_a|\tilde{i}\rangle=f_a\delta_{ab}+O(\frac{1}{N^2})$. The matrix elements here are taken with respect to the $d$ dimensional code words. Then the aKL equation can be written as 




\begin{equation}
\langle\tilde{j}|E_b^{\dagger}E_a|\tilde{i}\rangle=f_a\delta_{ab}\delta_{ij}+\frac{g_{ab}^{ij}}{N^2},
\end{equation}
where $f_a\geq 0$ is some $O(1)$ number that depends on the error indices.

Now we define two classes of errors according to whether $f_a$ is 0 or not.  \textit{The first class of errors}, labeled by $E_{\alpha}$ and $E_{\beta}$, satisfies $f_{\alpha}, f_{\beta}\neq 0$. \textit{The second class of errors}, labeled by $E_{\sigma}$ and $E_{\nu}$, satisfies $f_{\sigma}=f_{\nu}=0$.  Note that we will label the errors with Greek indices, $\alpha,\beta$ for the first class, $\sigma,\nu$ for the second class, when we distinguish their classes. When we do not need to treat them in different classes, we use Latin indices like $a,b$ to label these errors. For the first class of errors, we can define an orthonormal basis $\{|\mu_{\alpha}^i\rangle\}$ by acting error operators $E_{\alpha}$ on the logical states $|\tilde{i}\rangle$ such that

\begin{equation}
\begin{split}
    &E_{\alpha}|\tilde{i}\rangle =\sqrt{f_{\alpha}}|\mu_{\alpha}^i\rangle +\sum_{\beta} \frac{c_{\alpha\beta}}{2\sqrt{f_{\beta}}N^2}|\mu_{\beta}^i\rangle+\sum_{\beta} \frac{{\Delta}_{\alpha\beta}^i}{2\sqrt{f_{\beta}}N^2}|\mu_{\beta}^{i}\rangle +\sum_{\beta,j\neq i}\frac{h_{\alpha\beta}^{ij}}{2\sqrt{f_{\beta}}N^2}|\mu_{\beta}^j\rangle. \\
\end{split}
\end{equation}
The states $|\mu_{\alpha}^i\rangle$'s can be read off from the $O(1)$ term in above equations. All the other coefficients can be read off from the aKL equations as,
\begin{equation}\label{errorcoefficient}
\begin{split}
   & c_{\alpha\beta}=\frac{1}{d}\sum_{i=1}^d g_{\alpha\beta}^{ii}\\
    &\Delta^i_{\alpha\beta}=g_{\alpha\beta}^{ii}-c_{\alpha\beta}\\
    &h_{\alpha\beta}^{ij}=\begin{cases}
    g_{\alpha\beta}^{ij},&\text{if $i\neq j$ }\\
    0, &\text{if $i=j$}
    \end{cases}.
\end{split}
\end{equation}
One can see that there is a clear separation between the phase errors that are captured by $\Delta^i_{\alpha\beta}$, and the bit flip errors that are captured by $h_{\alpha\beta}^{ij}$. So far we have discussed the first class of errors with non-zero $O(1)$ norm $\sqrt{f_{\alpha}}$. However,   the second class of error operators have vanishing $O(1)$ norm and let us label them by $E_{\sigma}$ and $E_{\nu}$.  so $\langle E_{\sigma}^{\dagger}E_{\sigma}\rangle=O(\frac{1}{N^2})$. 
Generally, the aKL equations for these operators take the form of 
\begin{equation}\label{eqapseconderrormatrix}
    \langle\tilde{j}|E_{\sigma}^{\dagger}E_{\nu}|\tilde{i}\rangle=\frac{g_{\sigma\nu}^{ij}}{N^2}+O(\frac{1}{N^3}).
\end{equation}
These operators in the second class of errors map the logical states to states with norms proportional to $O(\frac{1}{N})$. More generally, they are operators that satisfy the following equation:
\begin{equation}\label{eqseccl}
    E_{\sigma}|\tilde{i}\rangle=\sum_{\nu}\frac{s_{\sigma \nu}^{ i}}{N}|\mu_{\nu}^i\rangle,
\end{equation}
for some $O(1)$ coefficients $s$'s. The states $|\mu_{\nu}^i\rangle$'s are another set of normalized basis, orthogonal to the states $|\mu_{\alpha}^i\rangle$'s up to $O(\frac{1}{N})$ overlap.   
The errors that satisfy
\begin{equation}
    \langle\tilde{i}|E_{\sigma}|\tilde{j\rangle}=\frac{e^{ij}_{\sigma}}{N}+O(\frac{1}{N^3}),
\end{equation}
 also belong to the second class of errors, since they are consistent with Eq.~\eqref{eqseccl}.

For some calculations, we do not need to differentiate the two types of errors. In that case, we use Latin indices and simply label all of them by $E_a$. Let us denote the KL matrix element by $\langle\tilde{j}|E_b^{\dagger}E_a|\tilde{i}\rangle=G_{ab}^{ij}$; one can write down a general relation
\begin{equation}
\begin{split}
    &E_a |\tilde{i}\rangle=\sum_{A}u_{aA}^i|\mu_{A}\rangle,
\end{split}
\end{equation}
where $\{|\mu_A\rangle\}$ is a set of orthonormal states, and $u^i_{aA}$ are coefficients that satisfy 
\begin{equation}
    \sum_{A} u_{a A}^iu_{b A}^j=G_{ab}^{ij}.
\end{equation}

\subsection{Early time approximation}\label{secapearlytime}
Recall that we model our open system dynamics as a Markovian process. At early time, we approximate the evolution by only including the first order perturbation of errors.\footnote{The coupling $\lambda_a(\epsilon_a)$ for each error operator $E_a$ is defined as $\lambda_a\gamma(\epsilon_a)$,, with $\gamma(\epsilon_a)$ defined in Eq.\ref{eqthermalfactor}. }

\begin{equation}
    \rho(t)\approx \rho_0-\sum_a\frac{\lambda_a(\epsilon_a) t}{2}\{E_a^{\dagger}E_a,\rho_0\}+\sum_a \lambda_a(\epsilon_a) t E_a\rho_0 E_a^{\dagger}+\cdots,
\end{equation}
where the sum over $E_a$ includes both the first and second class of errors, and $\rho_0$ is the maximally entangled state between reference $R$ and system $S$ at $t=0$,
\begin{equation}
    \rho_0=\frac{1}{d}\sum_{ij} |\tilde{i}i\rangle_{RS}\langle\tilde{j}j|.
\end{equation}


We proceed to compute the mutual information $I(R:S)=S(\rho^R)+S(\rho^S)-S(\rho)$ at early times. Since the non-trivial unitary dynamics that induces the above Markovian evolution only occurs between the system and the bath, the entropy of $R$ should remain invariant over time, \textit{i.e.} $S(\rho^R)=\ln d$. Therefore, it suffices to calculate the entropies of $S$ and the $RS$ joint system.

 Generally, the density matrix is block-diagonal. We can generate each block matrix using the operators $E_a$, with $a\in \mathscr{B}_n$, such that for any $a,b \in \mathscr{B}_n$, $G_{ab}\neq 0$. Here $\mathscr{B}_n$ is an index set that captures the error operators used to generate a particular block matrix. Sometimes we will also refer to this set as a block, since it defines the corresponding block matrix. We write the density matrix associated to the block $\mathscr{B}_n$ as,
\begin{equation}
\begin{split}
   \rho_n=& \sum_{a\in \mathscr{B}_n} E_{a}\rho_0 E^{ \dagger}_{a}\\
   =&\frac{1}{d}\sum_{a ABij} u_{a A}^iu_{a B}^{j *} |\mu_{A}\rangle _S\langle \mu_{B}|\otimes |i\rangle _R\langle j|.
\end{split}
\end{equation}

The eigenvalues of matrix $\frac{1}{d}\sum_{a}u_{a A}^i u_{a B}^{j *}$ is the same as that of the matrix $\frac{1}{d}\sum_{i A}u_{a A}^iu_{b A}^{i*}=\frac{1}{d}\sum_i G_{ab}^{ii}$.
After tracing out the reference $R$, the density matrix of system becomes 
\begin{equation}
    \rho_n^{S}=\frac{1}{d}\sum_{aiAB}u_{a A}^iu_{a B}^{i*}|\mu_{A}\rangle _S\langle \mu_{B}|.
\end{equation}
The eigenvalues of matrix $\frac{1}{d}\sum_{ai}u_{a A}^i u_{a B}^{i*}$ is the same as that of the matrix $\frac{1}{d}\sum_{ A}u_{a A}^iu_{b A}^{j*}= \frac{1}{d}G_{ab}^{ij}$.

Therefore, to evaluate the entropy, we need to calculate the eigenvalues of the matrix $\frac{1}{d}G_{ab}^{ij}$.  Generally, inside a block, we have both the first and the second class of errors. Their overlap $G_{\alpha\sigma}$ scales as $\frac{1}{N^2}$ by the aKL condition. 
However, we can always diagonalize the matrix $G$, where the diagonal form $\tilde{G}$ has vanishing matrix elements for the overlaps $\tilde{G}_{\alpha\sigma}=0$ and $\tilde{G}_{\alpha\beta}-G_{\alpha\beta}=O(\frac{1}{N^4}),\tilde{G}_{\sigma\nu}-G_{\sigma\nu}=O(\frac{1}{N^4})$ such that the correction is at subleading order. Therefore, in the following, we can neglect the overlap between the first and the second class of errors, and always treat them as different blocks. 

For our following discussion, we use $\rho_n^{(1)},\rho_n^{(2)}$ to denote the state created purely by the first and the second class of errors, respectively. The entropy for $\rho_n^{(1)}$ is,



\begin{equation}
\begin{split}
S(\rho_n^{(1)})&=-\Tr(\rho_n^{(1)} \ln\rho_n^{(1)})\\
&=-\Tr[(f+\frac{c}{N^2})\ln(f+\frac{c}{N^2})]\\
&=-\Tr(f\ln f)-\frac{1}{N^2}\Tr(c\ln f)-\frac{1}{N^2}\Tr(c)-\frac{1}{2N^4}\Tr(cf^{-1}c)+O(\frac{1}{N^6}).
\end{split}
\end{equation}

In the second step we used that the eigenvalues of the density matrix are the same as those of the KL overlap matrix $\frac{1}{d}\sum_iG^{ii}$. $f$ is the diagonal matrix $f_{\alpha}\delta_{\alpha\beta}$ and the matrix elements of $c$ are given by Eq.~\eqref{errorcoefficient}. 
For the reduced density matrix of system $S$, $\rho^S_n=\Tr_R(\rho_n)$, its entropy is 
\begin{equation}
\begin{split}
    S(\rho_n^{(1)S })=&-\Tr[\frac{1}{d}(f\otimes I+\frac{g}{N^2})\ln(f\otimes I+\frac{g}{N^2})]+\frac{1}{d}\Tr(f\otimes I+\frac{g}{N^2})\ln{d}\\
    =&\Tr(\rho_n)\ln{d}-\frac{1}{d}\Tr[(f\otimes I)\ln( f\otimes I)]-\frac{1}{dN^2}\Tr[g\ln (f\otimes I)]-\frac{1}{dN^2}\Tr(g)\\
    &-\frac{1}{2dN^4}\Tr[g(f^{-1}\otimes I)g]+O(\frac{1}{N^6}).
\end{split}
\end{equation}
 where $f$ is the matrix with element being $f_{\alpha\beta}$, and $I$ is the $d\times d$ identity matrix.    After subtracting $S(\rho_n^{(1)})$ from $S(\rho_n^{(1)S})$, we finally obtain 
\begin{equation}\label{eqapblockentropy}
\begin{split}
    S(\rho_n^{(1)S})-S(\rho_n^{(1)})&=\Tr(\rho_n^{(1)})\ln{d}-\frac{1}{2dN^4}\Tr[g(f^{-1}\otimes I )g]+\frac{1}{2N^4}\Tr(cf^{-1}c)+O(\frac{1}{N^6})\\
    &=\Tr(\rho_n^{(1)})\ln d-\frac{1}{2dN^4}(\sum_{i}\Tr[\Delta^i f^{-1}\Delta^i]+\sum_{ij}\Tr[h_{ij}f^{-1}h_{ij}^T])+O(\frac{1}{N^6}),
\end{split}
\end{equation}
where matrices $\Delta^i$ and $h_{ij}:=h^{ij}$ are defined in Eq.~\eqref{errorcoefficient}.  We see that the decay rate of mutual information is suppressed by $\frac{1}{N^4}$. Repeating the same subtractions but for the second class of errors, the difference in entropy is 


\begin{equation}\label{eqapsecondentropy}
\begin{split}
    S(\rho_n^{(2)S})-S(\rho_n^{(2)})=&\frac{1}{N^2}[S(\frac{1}{d}g^{ij}_{\sigma\nu})-S({\frac{1}{d}\sum_ig^{ii}_{\sigma\nu}})]|_{\sigma,\nu\in\mathscr{B}_n}+O(\frac{1}{N^4})\\
    =&\Tr(\rho_n^{(2)})\ln{d}-\frac{1}{N^2}[S(\frac{1}{d}{\sum_ig^{ii}_{\sigma\nu}})-\frac{1}{d}S(g^{ij}_{\sigma\nu})]|_{\sigma,\nu\in\mathscr{B}_n'}+O(\frac{1}{N^4}),
\end{split}
\end{equation}
where $S(g_{\sigma\nu}^{ij})$ means that we are calculating  entropy of the matrix whose elements are $g_{\sigma i,\nu j}$. On the other hand, in $S(\frac 1 d \sum_i g_{\sigma\nu}^{ii})$, we compute the entropy of the matrix with elements proportional to $\frac 1 d\sum_i g^{ii}_{\sigma\nu}$. The decay rate is only suppressed in $\frac{1}{N^2}$. The entropy difference in the second term depends on the matrix elements $g_{\sigma\nu}^{ij}$ in a complicated way. Instead of calculating it explicitly, we just provide an upper bound for this expression.  Suppose that in each block $\mathscr{B}_n$, the matrix $g_{\sigma\nu}^{ij}$ has a rank of $r(n)$ with its elements bounded by a function $f(n)$, then the second term is bounded by $\frac{r(n)f(n)\log{d}}{N^2}$. 


For $n\geq 1$, define $\alpha_n=\lambda_n t$. The early time density matrix can be written as\footnote{To keep the notation tractable, we absorbed any $a$-dependence of the coupling $\lambda_a$ into the operators $E_a$ such that the couplings $\lambda$ only retain the $n$ dependence. To be more precise, one can redefine the couplings and errors such that $\lambda(\epsilon_a)$ only depends on the energy of each error operator $\epsilon_a$. Because $\epsilon_a=\epsilon_n$ for all $a\in\mathscr{B}_n$, we can rewrite $\lambda(\epsilon_a)$ as $\lambda_n$. }

\begin{equation}
    \rho(t)=\bigoplus_{n=0}^{\infty} (\alpha_n    \rho_n).
\end{equation}

Applying the relation $\sum_n \alpha_n \Tr(\rho_n)=1$, one find that the difference of total entropy is 
\begin{equation}\label{eqapentropy}
\begin{split}
    S(\rho^S(t))-S(\rho(t))=&\ln{d}-\sum_{n=1}^{\infty}\frac{\lambda_nt}{N^2} [S({\frac{1}{d}\sum_ig^{ii}_{\sigma\nu}})-\frac 1 dS(g^{ij}_{\sigma\nu})]|_{\sigma,\nu\in\mathscr{B}_n'}\\ &-\sum_{n=1}^{\infty} \frac{\lambda_n t}{2dN^4}\sum_{\alpha\beta\in\mathscr{B}_n}[ \sum_i\Delta^{i2}_{\alpha\beta}+ \sum_{i\neq j}(h_{\alpha\beta}^{ij})^2 ]+O(\frac{1}{N^6}).
\end{split}
\end{equation}

For early time, i.e., $\lambda_n t\ll 1$, the dominant contribution to information decay is from the second class of errors. 

\subsection{Finite time }
In this section, we look at the finite time behavior of density matrix up to $O(\frac{1}{N^2})$ contributions. We will see that, although the early-time effect of the first class of errors is sub-dominant, it becomes more relevant at later times since it grows quadratically in time. Recall that the overlap between the different classes of errors are suppressed, we can split the Lindbladian operator into two parts associated with  the first class and the second class of errors, 

\begin{equation}
    \rho(t)=e^{\mathcal{L}t}\rho(0)=e^{\mathcal{L}^{(1)}t+\mathcal{L}^{(2)}t}\rho(0) + O(\frac{1}{N^4}).
\end{equation}

From Eq.~\eqref{eqseccl}, the second class of Lindblad operator creates a term of $O(\frac{1}{N^2})$ each time it acts on the density matrix. Therefore, the state can be approximated by 
\begin{equation}
    \rho(t)=\rho^{(1)}(t)+t\mathcal{L}^{(2)}[\rho^{(1)}(t)]+O(\frac{1}{N^4}),
\end{equation}
with $\rho^{(1)}(t)=e^{\mathcal{L}^{(1)}t}\rho(0)$ being the state evolved under the first class of errors. Here we only keep the commuting contributions from $\mathcal{L}^{(1)}$ and $\mathcal{L}^{(2)}$, since their commutator only contributes at higher orders of $\frac{1}{N}$.

To simplify the discussion, we will look at an approximation, which we claim is sufficient to capture the behavior of information decay in our discussion. Let us  first look at the case with only one type of error $E$ in the first class,

\begin{equation}
    \dot{\rho}(t)=\lambda E\rho  E^{\dagger}+\lambda E^{\dagger}\rho E-\frac{\lambda}{2}\{E^{\dagger}E,\rho\}-\frac{\lambda}{2}\{EE^{\dagger},\rho\}.
\end{equation}

For low energy operators $E$, we can make the approximation where the coupling of $E$ and its conjugate has the same strength $\lambda$, which in general should be different due to the thermal factor. In this large $N$ model, we observe that $E^{\dagger}E$ acts trivially in the leading order, 
\begin{equation}
\begin{split}
E^{\dagger}E E^p\rho &=(p+1)E^p\rho+O(\frac{1}{N^2})\\
EE^{\dagger}E^p\rho&=pE^p\rho +O(\frac{1}{N^2}).
\end{split}
\end{equation}
where $p$ is some integer power.

The order  $\frac{1}{N^2}$ correctons can be captured by an operator $Z$, which acts on the code state in the following way: 
\begin{equation}
    Z|\mu^i_{\alpha}\rangle =(1+\frac{g^{ii}}{2N^2})|\mu^i_{\alpha}\rangle.
\end{equation}

It is not hard to find the solution up to order $\frac{1}{N^2}$,
\begin{equation}
    \rho(t)=\sum_{k=0}^{\infty} \frac{(\lambda t)^k}{(1+\lambda Z t)^{k+1}}\frac{E^k\rho_0 E^{\dagger k}}{k!}+O(\frac{1}{N^4}).
\end{equation}


One might generalize this equation to include all the first class error operators $E_{\alpha}$. We denote the density matrix dressed by the first class of errors as $\rho^{(1)}(t)$, which is  

\begin{equation}\label{eqapdensitymatrix}
    \rho^{(1)}(t)=\sum_{\{k_{\alpha}\}}w(\{k_{\alpha}\})\prod_{\alpha} \frac{ E_{\alpha}^{k_{\alpha}}\rho(0) E_{\alpha}^{\dagger k_{\alpha}}}{ k_{\alpha}!}+O(\frac{1}{N^4}),
\end{equation}
where the weight function can be upper bounded by,

\begin{equation}\label{eqweigthbound}
\begin{split}
   & w(\{k_{\alpha}\})=\prod_{\alpha}w(k_{\alpha}),\\
  & \mathrm{where}~w(k_{\alpha})\leq\frac{(\lambda_{\alpha} t)^{k_{\alpha}}}{(1+\lambda_{\alpha} Z_{\alpha} t)^{k_{\alpha}+1}}, \ \ \text{for $k_{\alpha}\geq 1$}.
\end{split}
\end{equation}

The bound is only saturated if we could neglect the difference between the coupling constants of $E_{\alpha}$ and $E_{\alpha}^{\dagger}$. While said difference is negligible for low energy errors, it is not for high energy ones. In fact, the coupling constant goes down exponentially with the energy $\epsilon_{\alpha}$ of the operator $E_{\alpha}$,

\begin{equation}
\lambda_{\alpha}\propto\gamma(\epsilon_{\alpha})\sim e^{-\beta \epsilon_{\alpha}},
\end{equation}
for $\epsilon_{\alpha}>0$. However, the coupling constant $\lambda_{\alpha}'$ for $E_{\alpha}^{\dagger}$ barely depends on energy, since $\gamma(\epsilon_{\alpha})\sim 1$, for $\epsilon_{\alpha}<0$. We only bound the weight function for $k_{\alpha}\geq 1$, since the $k_{\alpha}=0$ component does not contribute to the mutual information decay, and it is determined by the constraint $\sum_{k_{\alpha}=0}^{\infty}w(k_{\alpha})=1$ of the weight function.

Given the density matrix, we calculate the entropy difference by the equation,

\begin{equation}
S(\rho^{(1)S})-S(\rho^{(1)})=\Tr(\rho^{(1)})\ln{2}-\frac{1}{2dN^4}\sum_{\mathcal{A}\mathcal{B}} w_{\mathcal{A}}(\sum_i(\Delta_{\mathcal{A}\mathcal{B}}^i)^2+\sum_{i\neq j}(h_{\mathcal{A}\mathcal{B}}^{ij})^2)+O(\frac{1}{N^6}),
\end{equation}
where we use $\mathcal{A}$ and $\mathcal{B}$ as a shorthand notation for the product of first class operators, so that $E_{\mathcal{A}}:=\prod_{\alpha_i\in \mathcal{A}}E_{\alpha_i}$. $w_{\mathcal{A}}$ is a number obtained by setting all the $Z_{\alpha}$'s to identity in Eq.~\eqref{eqweigthbound}.  The $d$-function and $h$-function scale linearly with $k_n$, and are proportional to some power $q$ of $n_{\alpha}$, with $n_{\alpha}=\frac{\epsilon_{\alpha}}{\Omega}$ being the energy level of $E_{\alpha}$, where $\Omega$ has dimension of energy. For example, $\Omega=\omega$ for our oscillator model.  Thus we can upper bound the information decay rate by (up to $\frac{1}{N^4}$),

\begin{equation}\label{decayratebound}
\begin{split}
K^{(1)}(t)&=\frac{1}{N^4}\sum_{k_{\alpha}=1}^{\infty}w(\{k_{\alpha}\})(\sum_{\alpha} n_{\alpha}^qk_{\alpha})^2\\
&\leq \frac{1}{N^4}[ 2(\sum_{\alpha} \lambda_{\alpha} n_{\alpha}^q t)^2+\sum_{\alpha}\lambda_{\alpha} n_{\alpha}^{2q} t].
\end{split}
\end{equation}

Note that the coefficient $\lambda_{\alpha}$ scales as $e^{-\beta\epsilon_{\alpha}}$, and the density of errors with energy $\epsilon$ is bounded by $e^{\mu\epsilon}$ by our sparsity condition. By the same argument we have for Eq.(\ref{eqn:hagedorn}) in Section~\ref{case1}, one can show that the quantity inside the square bracket does not scale as $N$ as long as $T<\frac{1}{\mu}$.






The decay rate due to the first class of errors is suppressed by $N^4$ and grows quadratically with time for $\lambda t\gg 1$.  Similar arguments also apply for higher orders of $\frac{1}{N^2}$. In general, the decay rate at order $\frac{1}{N^{2m}}$ is proportional to $\Delta^m$ and $h^m$ (defined in Eq.~\eqref{errorcoefficient}) of the corresponding error operators, which scales as $k^m$ for the error operator $E_{\alpha}^k$. Therefore, the general decay rate takes the form of $\frac{t^m}{N^{2m}}$ at each order with $m\geq 2$.
 
 The second class of errors of the form $E_{\nu}^k$  is suppressed by $\frac{t^k}{N^{2k}}$ in the Lindblad equation, where the leading order term is linear in $t$, 
 \begin{equation}
 \begin{split}
  \rho^{(2)}(t)&:=t\mathcal{L}^{(2)}[\rho^{(1)}(t)] \\
  &= \sum_{\nu} \lambda_{\nu}t[ E_{\nu}\rho^{(1)}(t)E_{\nu}^{\dagger}-\frac{1}{2}\{\rho^{(1)}(t),E_{\nu}^{\dagger}E_{\nu}\}].
  \end{split}
 \end{equation}

Recall that $\rho^{(1)}(t)$ is the density matrix induced by the first class of errors, explicitly given by Eq.~\eqref{eqapdensitymatrix}. For the sake of clarity, let us rewrite $\rho^{(1)}(t)$ as
\begin{equation}
    \rho^{(1)}(t)=\sum_{\mathcal{A}} w_{\mathcal{A}}(t) E_{\mathcal{A}}\rho_0E_{\mathcal{A}}^{\dagger}.
\end{equation}
Like our discussion in Appendix~\ref{secapearlytime}, we convert the entropy of the state to the entropy of the matrix  $g_{\sigma\mathcal{A},\nu\mathcal{B}}^{ij}$, which can be read off from
\begin{equation}
    \langle\tilde{j}| E^{\dagger}_{\mathcal{B}}E^{\dagger}_{\nu}E_{\sigma}E_{\mathcal{A}}|\tilde{i}\rangle=\frac{g_{\mathcal{A}\sigma,\mathcal{B}\nu}^{ij}}{N^2}+O(\frac{1}{N^4}).
\end{equation}

Being the product of a second class error and a first class error, this composite operator belongs to the second class of errors, but it has  $O(\frac{1}{N})$ overlap with some operators in the first class, 
\begin{equation}\label{eqapoverlap}
    \langle\tilde{j}|E_{\mathcal{B}}^{\dagger}E_{\sigma}E_{\mathcal{A}}|\tilde{i}\rangle=\frac{\delta_{ij}g_{\sigma\mathcal{A},\mathcal{B}}}{N}+O(\frac{1}{N^3}).
\end{equation}
Due to the factorization property, $\langle\tilde{j}| E^{\dagger}_{\mathcal{B}}E^{\dagger}_{\nu}E_{\sigma}E_{\mathcal{A}}|\tilde{i}\rangle=\langle\tilde{j}|E_{\nu}^{\dagger}E_{\sigma}|\tilde{i}\rangle\langle0| E_{\mathcal{B}}^{\dagger}E_{\mathcal{A}}|0\rangle+\langle\tilde{j}|\tilde{i}\rangle\langle0| E_{\mathcal{B}}^{\dagger}E_{\nu}^{\dagger}E_{\sigma}E_{\mathcal{A}}|0\rangle$, the matrix element can be written as sum of two terms, 
\begin{equation}\label{eqapcomposite}
    g^{ij}_{\mathcal{A}\sigma,\mathcal{B}\nu}=g^{ij}_{\sigma\nu}f_{\mathcal{A}\mathcal{B}}+\delta^{ij}g_{\sigma\mathcal{A},\nu\mathcal{B}}.
\end{equation}
where $f_{\mathcal{A}\mathcal{B}}:=\langle 0| E_{\mathcal{B}}^{\dagger}E_{\mathcal{A}}|0\rangle$, and $|0\rangle$ is the vacuum state of our system. The matrix $g^{ij}_{\sigma\nu}$ is identical to the one in Eq.~\eqref{eqapseconderrormatrix}.  $g_{\sigma\mathcal{A},\nu\mathcal{B}}$ is positive definite where each element is equal to the  four point function $\langle 0| E_{\mathcal{B}}^{\dagger}E_{\nu}^{\dagger}E_{\sigma}E_{\mathcal{A}}|0\rangle $, modulo the $\frac{1}{N^2}$ factor.  The overlap between the composite operators and some first class operators (Eq.~\eqref{eqapoverlap}) can be removed by a basis transformation, which also cancels the second term in Eq.~\eqref{eqapcomposite}. 
Therefore, similar to Eq.~\eqref{eqapsecondentropy}, we conclude that the entropy difference is
\begin{equation}
\begin{split}
    S(\rho^{(2) S}(t))-S(\rho^{(2)}(t))=&\frac{1}{N^2}[S(\frac{1}{d}g^{ij}_{\sigma\nu}(t)f_{\mathcal{A}\mathcal{B}}(t))-S(\frac 1 d\sum_i g^{ii}_{\sigma\nu}(t)f_{\mathcal{A}\mathcal{B}}(t))]\\
    =&\frac{\Tr(f_{\mathcal{A}\mathcal{B}}(t))}{N^2}[S(\frac 1 dg_{\sigma\nu}^{ij}(t))-S(\frac 1 d\sum_ig_{\sigma\nu}^{ii}(t))]\\
    =&\sum_n \frac{\lambda_n t}{N^2}[ S(\frac 1 dg_{\sigma\nu}^{ij})-S(\frac 1 d\sum_ig_{\sigma\nu}^{ii})]|_{\sigma,\nu\in \mathscr{B}_n'},
\end{split}
\end{equation}
where $S(\frac 1 dg_{\sigma\nu}^{ij}(t)f_{\mathcal{A}\mathcal{B}}(t))$ denotes the entropy of the matrix whose element is $\frac 1 dg_{\sigma\mathcal{A}i,\nu\mathcal{B}j}(t)=\frac 1 dg_{\sigma\nu}^{ij}(t)f_{\mathcal{A}\mathcal{B}}(t)$. We have absorbed the factor $\lambda_{\nu}t$ into the matrix $g_{\sigma\nu}^{ij}(t):=\sqrt{\lambda_{\nu}\lambda_{\sigma}} t g_{\sigma\nu}^{ij}$, and the weight function $w_{\mathcal{A}}(t)$ into the function $f_{\mathcal{A}\mathcal{B}}(t):=\sqrt{w_{\mathcal{A}}(t)w_{\mathcal{B}}(t)}f_{\mathcal{A}\mathcal{B}}$. By definition, $\Tr(f_{\mathcal{A}\mathcal{B}})=\sum_{\mathcal{A}}w_{\mathcal{A}}(t)\langle 0|E_{\mathcal{A}}^{\dagger}E_{\mathcal{A}}|0\rangle=\Tr(\rho^{(1)}(t))+O(\frac{1}{N^2})=1+O(\frac{1}{N^2})$. In the last step, we again utilized the block diagonal form of the matrix $g_{\sigma\nu}^{ij}$ to write the result as a sum of each block's contribution, with blocks being labeled by $\mathscr{B}_n'$,  such that the matrix elements are non-zero only for $\sigma$ and $\nu$ in the same block. Note that the entropy difference is the same as in Eq.~\eqref{eqapsecondentropy}. We can bound the decay rate induced by the second class of errors with the following function at leading order of $\frac{\lambda t}{N^2}$,


\begin{equation}
    K^{(2)}(t)\leq\sum_n  \frac{\lambda_n t}{N^2}f(n)r(n)\log{d},
\end{equation}
where $f(n)$ is a upper bound for each matrix element $g_{\sigma\nu}^{ij}$ in the block $\mathscr{B}_n$, and $r(n)$ is the rank of the block.

At higher order, we can continue to expand the second class of Lindbladian operator and one can repeat these calculations following the same procedure. Errors of the form $E_{\sigma}^m$ then induce information decay at $O(\frac{1}{N^{2m}})$, with magnitude proportional to $(\lambda t)^m$. 

In summary,  we find that both the first class of errors and the second class of errors contribute to the information decay with some rate function  $K_m(t)\propto \frac{t^m}{N^{2m}}$ as long as the assumptions of Theorem~\ref{thm} are satisfied. For $N^2\gg\lambda t\gg 1$,


\begin{equation}
    I(S:R)=2\ln{d}-\sum_{m=1}^{\infty}K_m(t).
\end{equation}
 

\section{aKL condition for single trace errors}
\subsection{non-local model}\label{apphase}
In this section, we present the aKL equations for the most general single-trace error operators.  Since the logical states (Eq.\ref{eqnlls2}) are symmetric in $a_1^{\dagger}$ and $a_2^{\dagger}$, we only need to consider the single-trace operators involving $a_1$ and $a_1^{\dagger}$. In the following, we will only keep track of the terms up to
 $O(\frac{1}{N^2})$.

Let us define a class of error operators

\begin{equation}
    E_{n,m}^P=\frac{:\Tr(P\{a_1^{\dagger n} a_1^m\}):}{N^{\frac{n+m}{2}}},
\end{equation} 
where $P$ denotes a particular permutation of the operators inside the curly brackets. 
Acting it on the logical states, one obtains,
\begin{equation}\label{apB1}
    E_{n,m}^P|\tilde{i}\rangle=f|\mu^i\rangle+\frac{g}{N^2}|\mu\rangle+O(\frac{1}{N^4}).
\end{equation}

To extract these coefficients, we calculate the overlap
\begin{equation}
    \langle\tilde{i}|E_{n',m'}^{P'\dagger}E_{n,m}^P|\tilde{j}\rangle,
\end{equation}
which is essentially a four-point function that factorizes into,  
\begin{equation}
    \langle \tilde{O}_i^{\dagger} E^{P'\dagger}_{n',m'}E_{n,m}^P\tilde{O}_j\rangle=\langle\tilde{O_i}^{\dagger}\tilde{O_j}\rangle \langle E^{P'\dagger}_{n',m'}E_{n,m}^P\rangle+O(\frac{1}{N}).
\end{equation}

This is the only way to factorize into disconnected correlators, because $O_i$ and $O_j$ contain $a_2^{\dagger}$, while $E_{n,m}^P$'s do not. The two point function $\langle E^{P'\dagger}_{n'm'}E_{n,m}^P\rangle$ vanishes, unless $m,m'=0$, which is the case that we discussed in the main text. Moreover, the sub-leading term can be $O(\frac{1}{N})$ only if $m'=n'=0$, hence the only non-trivial four-point function contribution is when $m=n$.
In this case, we define $Z_n^P=\frac{:\Tr(P\{a^{\dagger n}a^n\}):}{N^n}$ and find that,
\begin{equation}\label{eqn:errDet}
    \langle \tilde{j}|Z_n^P|\tilde{i}\rangle =\frac{c_n^P}{N}+O(\frac{1}{N^3}), 
\end{equation}
for some constant $c_n^P$, which is non-zero only when $n=2$. This is consistent with the aKL condition (Eq.~\eqref{eqknill}).



So far, we have discussed the order one and $\frac{1}{N}$ term in Eq.~\eqref{eqn:errDet}. All the other error operators can only produce order $\frac{1}{N^2}$ terms when acting on the logical states. This completes the proof of aKL condition.

As a special example, consider the decay mode,  $E_{0,2}$, which lower the energy when acting on the logical states. It also satisfies the aKL condition, 
\begin{equation}
\begin{split}
    \langle\tilde{1}|E_{0,2}^{\dagger}E_{0,2}|\tilde{1}\rangle-\langle\tilde{2}|E_{0,2}^{\dagger}E_{0,2}|\tilde{2}\rangle=\frac{8}{N^2}+O(\frac{1}{N^4})\\
    \langle\tilde{2}|E_{0,2}^{\dagger}E_{0,2}|\tilde{1}\rangle=\frac{4\sqrt{2}}{N^3}+O(\frac{1}{N^5}).
\end{split}
\end{equation}

So the decay mode belongs to the second class of errors classified in Appendix~\ref{apentropy}, and lead to a spontaneous decay of quantum information at zero temperature with a rate proportional to  $\frac{1}{N^2}$.

\subsection{Local model}\label{aploaKL}
In this section, we prove the aKL condition for the single-trace errors of the form 

\begin{equation}
E_{\{n_i\}}=\frac{:\Tr(P\{a_1^{\dagger n_1}a_2^{\dagger n_2}a_1^{n_3}a_2^{n_4}\}):}{N^{\frac{\sum_{k=1}^4 n_k}{2}}}
\end{equation}
with logical states defined as 
\begin{equation}
\begin{split}
    |\tilde{\uparrow}\rangle&=\frac{\Tr(a_1^{\dagger})}{\sqrt{N}}|0\rangle_{12}\\
    |\tilde{\downarrow}\rangle&=\frac{\Tr(a_2^{\dagger})}{\sqrt{N}}|0\rangle_{12} .
\end{split}
\end{equation}

In the main text, we derived a general formula for the case when $n_3=n_4=0$. Moreover, since the logical states are annihilated by more than two $a_1$ or $a_2$, and the error operators are normal ordered, we are left to discuss the case when $n_3+n_4=1$. Without loss of generality, we will only consider the case with $n_3=1,\ n_4=0$. Then the relevant overlaps are, 
\begin{equation}
\begin{split}
   & \langle\tilde\uparrow|E^{\dagger}_{(n-1,m,0,0)}E_{(n,m,1,0)}|\tilde\uparrow\rangle=\frac{n}{N^2}+O(\frac{1}{N^4})\\
   &\langle\tilde\uparrow|E^{\dagger}_{(n,0,1,0)}E_{(n,0,1,0)}|\tilde\uparrow\rangle=\frac{n}{N^2}+O(\frac{1}{N^4})\\
   & \langle\tilde\downarrow|E^{\dagger}_{(n,m,0,0)}E_{(n,m,1,0)}|\tilde\downarrow\rangle=\langle\tilde\downarrow|E^{\dagger}_{(n,0,1,0)}E_{(n,0,1,0)}|\tilde\downarrow\rangle=0\\
 &  \langle\tilde\downarrow|E^{\dagger}_{(n,m-1,0,0)}E_{(n,m,1,0)}|\tilde\uparrow\rangle=\frac{m}{N^2}+O(\frac{1}{N^4}).
\end{split}
\end{equation}

Exceptions of these formulae consist of special cases when $n=1,\ m=0$ and $n=0,m=1$, where one can derive that, 
\begin{equation}
\begin{split}
  & \langle\tilde\uparrow|\frac{\Tr(a_1^{\dagger}a_1)}{N} |\tilde\uparrow\rangle=\frac{1}{N}, \ \ \langle\tilde\downarrow|\frac{\Tr(a_1^{\dagger}a_1)}{N} |\tilde\downarrow\rangle=0\\
   &\langle\tilde\downarrow|\frac{\Tr(a_2^{\dagger}a_1)}{N} |\tilde\uparrow\rangle=\frac{1}{N}.
 \end{split}
\end{equation}
These are $O(\frac{1}{N})$ overlap, which are consistnet with Eq.~\eqref{eqknill}. So we complete the proof of aKL~(\ref{eqknill}) for general single-trace errors in this local model. 




\subsection{Global symmetry}\label{globaleg}

In this section, we prove the aKL condition  for the model in Section~\ref{case3}. The logical space is spanned by the two states of unit norm
\begin{equation}
    \begin{split}
        &|1\rangle=\frac{\Tr(a^{\dagger L})^2}{N^L\xi(L,L)}\\
        &|2\rangle=\frac{\Tr(a^{\dagger L+1})\Tr(a^{\dagger L-1})}{N^L\xi(L+1,L-1)},
    \end{split}
\end{equation}
where the normalization constant are $\xi(L,L)=\sqrt{2}L(1+O(\frac{1}{N^2}))$ and $\xi(L+1,L-1)=\sqrt{L^2-1}(1+O(\frac{1}{N^2}))$. Note that these states are not orthogonal but have overlap 
\begin{equation}
    \langle1|2\rangle=\frac{L^2\sqrt{L^2-1}}{\sqrt{2}N^2}.
\end{equation}

To check the aKL  condition, we insert the error operators  $E_n=\frac{\Tr(a^{\dagger n})}{N^{\frac{n}{2}}\xi(n)}$ with $\xi(n)$ being the normalization constant, and derive the following equations

\begin{equation}\label{KLglobal}
\begin{split}
    &\langle 1|E_m^{\dagger}E_n|1\rangle-\langle 2|E_m^{\dagger}E_n|2\rangle =\delta_{mn}[\frac{2n}{N^2}+O(\frac{1}{N^4})]\\
    &\langle 1|E_m^{\dagger}E_n|2\rangle=\delta_{mn}[\frac{L^2\sqrt{L^2-1}}{\sqrt{2}N^2}+O(\frac{1}{N^4})].
\end{split}
\end{equation}

Note that the leading term in the second equation is a consequence of the non-orthogonality between our definition of $|1\rangle$ and $|2\rangle$; it can be easily removed via a change of basis (c.f. equation ~\ref{eqglbasis}). The above equation also holds for $E_n=\frac{\Tr(a^{ n})}{N^{\frac{n}{2}}\xi(n)}$.

For more general operators $E_{(n,m)}:=\frac{\Tr(a^{\dagger n}a^m)}{N^{\frac{m+n}{2}}}$, with $n\neq m$, one can show that 
\begin{equation}
\begin{split}
    &\langle 1|E_{n',m'}^{\dagger}E_{n,m}|1\rangle-\langle 2|E_{n',m'}^{\dagger}E_{n,m}|2\rangle =\delta_{n'-m',n-m}[\frac{2}{N^2}+O(\frac{1}{N^4})]\\
    &\langle 1|E_{n',m'}^{\dagger}E_{n,m}|2\rangle=\delta_{n'-m',n-m}[\frac{L^2\sqrt{L^2-1}}{\sqrt{2}N^2}+O(\frac{1}{N^4})],
\end{split}
\end{equation}
which indicates the aKL condition holds.

For the special case of $n=m$, the operator also incurs no phase error, because
\begin{equation}
   \langle 1|\frac{\Tr(a^{\dagger n}a^n)}{N^n} |1\rangle=\langle 2|\frac{\Tr(a^{\dagger n}a^n)}{N^n} |2\rangle=\frac{2L}{N}.
\end{equation}

We thereby conclude that the aKL conditions hold for all the single trace operators.

\subsection{Spin model}\label{apspin}
The approximate Knill-Laflamme condition for the spin model can be calculated using the spin algebra. Recall that  we can construct two states that span the code subspace in a way analogous to our definition in the oscillator model, 
\begin{equation}
\begin{split}
&|1\rangle=\frac{\Tr(S^{+ L})^2}{N^L\eta(L,L)}\\
&|2\rangle=\frac{\Tr(S^{+ L-1})\Tr(S^{+ L+1})}{N^L\eta(L+1,L-1)},
\end{split}
\end{equation}
where $\eta(L,L')$ is the normalization constant. They have the same leading order behavior as $\xi(L,L')$ in the harmonic oscillator model, but have different forms in sub-leading terms.  The overlap of these two states is 
\begin{equation}
\langle 1|2 \rangle =\frac{f(L)L\sqrt{L^2-1}}{\sqrt{2}N^2},
\end{equation}
where $f(L)$ is some complicated function of $L$. Again, we insert error operators of the form $E_n=\frac{\Tr(S^{\pm n})}{N^{\frac{n}{2}}\eta(n)}$ and $E_{n,m}=\frac{\Tr(S^{+n}S^{-m})}{N^{\frac{m+n}{2}}}$ into the correlator, and derive the aKL equations for the spin model,
\begin{equation}\label{KLglobal}
\begin{split}
    &\langle 1|E_m^{\dagger}E_n|1\rangle-\langle 2|E_m^{\dagger}E_n|2\rangle =\delta_{mn}[\frac{2n}{N^2}+O(\frac{1}{N^4})]\\
    &\langle 1|E_{n',m'}^{\dagger}E_{n,m}|1\rangle-\langle 2|E_{n',m'}^{\dagger}E_{n,m}|2\rangle =\delta_{n'-m',n-m}[\frac{2}{N^2}+O(\frac{1}{N^4})].
\end{split}
\end{equation}

Again, the bit flip error only appears at $O(\frac{1}{N^4})$

\begin{equation}
     \langle 1|E_m^{\dagger}E_n|2\rangle=\delta_{mn}[\frac{f(L)L\sqrt{L^2-1}}{\sqrt{2}N^2}+O(\frac{1}{N^4})],
\end{equation}
where we recall the first term can be removed by choosing an orthonormal basis. This completes our proof for the aKL condition.

\section{ Master equation in Lindblad form} \label{apLind}
Suppose the coupling between system and thermal reservoir is 
\begin{equation}
\begin{split}
    H=&H_S+H_B+V_I\\
    V_I=&\sum_a E_a\otimes b_a+h.c.,
\end{split}
\end{equation}
where $H_S$ and $H_B$ are the system and bath Hamiltonian respectively, and $b_a$'s are the bath modes. 

In the interaction picture, time evolution is described by 
\begin{equation}
\begin{split}
    \dot{\rho}(t)=&-i[V_I(t),\rho(t)]\\
    =&-i[V_I(t),\rho_0]-[V_I(t),\int_0^t dt'[V_I(t'),\rho(t')]].
\end{split}
\end{equation}

The essential simplification is enabled by the Born-Markov approximation, which allows one to approximate the full density matrix as $\rho(t)\sim \rho_S(t)\otimes\rho_B$ at all times.  After tracing out the thermal state, we arrive at, 
 \begin{equation}\label{eqn:thermtrace}
    \dot{ \rho}_{IS}(t)=\int_0^t dt'\sum_{ab} [E_b(t')\rho_{IS}(t)E_a^{\dagger}(t)  -E_a^{\dagger}(t)E_b(t')\rho_{IS}(t)]\Tr(\rho_Bb_a^{\dagger}(t)b_b(t'))+h.c. .
 \end{equation}
 
For sufficiently long time $t$, we can treat the initial time as $-\infty$. We also assume that the bath modes are uncorrelated. Then we define
 \begin{equation}
    \int_{-\infty}^t dt' \Tr(\rho_B b_a^{\dagger}(t)b_b(t'))e^{i\omega(t-t')}:=\frac{\gamma(\omega)}{2}\delta_{ab}.
 \end{equation}
 
For  operators $E_a(t)$ with the decomposition,
 \begin{equation}
     E_a(t)=\sum_kE_a^{(k)}e^{-i\epsilon_a^{(k)}t},
 \end{equation}
Eq (\ref{eqn:thermtrace}) simplifies as, 
 \begin{equation}
    \dot{ \rho}_{IS}(t)=\sum_{a,k} \frac{\gamma(\epsilon_a^{(k)})}{2}[E_a^{(k)}\rho_{IS}(t)E_a^{\dagger}(t) -E_a^{\dagger}(t)E_a^{(k)}\rho_{IS}(t)]e^{-i\epsilon_a^{(k)}t} +h.c. .
 \end{equation}
 
For example, consider the error operator $E_{ijkl}=a^{\dagger i}_ja^{\dagger k}_l$ in Section~\ref{case3}. It can be decomposed as, 
 \begin{equation}
 \begin{split}
      E_{ijkl}(t)=&\widetilde{a^{\dagger i}_ja^{\dagger k}_l}e^{-2i\mathcal{J}t}+[\frac{\widetilde{(a^{\dagger 2})^i_l}}{N}\delta^k_j+\frac{\widetilde{(a^{\dagger 2})^k_j}}{N}\delta^i_l+\frac{\Tr(a^{\dagger})}{N}\tilde{a}^{\dagger i}_j\delta^k_l+\frac{\Tr(a^{\dagger})}{N}\tilde{a}^{\dagger k}_l\delta^i_j]e^{-i\mathcal{J}t}\\
    &+\frac{\Tr(a^{\dagger 2})}{N^2}\delta^i_l\delta^k_j+\frac{\Tr(a^{\dagger})^2}{N^2}\delta^i_j\delta^k_l,
    \end{split}
 \end{equation}
 where the operator under the tilde vanishes when contracting with the kronecker delta, e.g. $a^{\dagger i}_j\delta_i^j=0$. Consequently, the tensor coefficients are almost orthogonal to each other, \textit{i.e.},
 \begin{equation}\label{apeqoverlap}
     \sum_{ijkl}E_{ijkl}^{\dagger (m)}E_{ijkl}^{(n)}\propto \delta^{mn}+O(\frac{1}{N}),
 \end{equation}
where the $O(\frac{1}{N})$ term denotes operators that has $O(\frac{1}{N})$ expectation value with respect to any low energy state.

More generally,
a tensor product of individual creation and annihilation operators can always be decomposed into different components obtained by index contractions, which are orthogonal to each other up to $O(\frac{1}{N})$, in the sense of Eq.~\eqref{apeqoverlap}.
As a result, the Lindbladian operator can always be written as a diagonal sum of components that correspond to different contractions,
\begin{equation}
   \sum_m \gamma(\epsilon^{(m)}) \sum_{a}E_{a}^{(m)}\rho E^{\dagger (m)}_{a}=\sum_{m a} \gamma(\epsilon^{(m)})E_a^{(m)}\rho E_a^{\dagger(m)}.
\end{equation}

\section{Expectation value of the Casimir}\label{apcasimir}
In this section, we  evaluate the expectation value of the penalty Hamiltonian $H_G$ (in Eq.~\eqref{eqcasi}) for a thermal state $\rho_T$ at temperature $T<\frac{\mathcal{J}}{2\log{N}}$,
\begin{equation}
    \Tr(\rho_T H_G)=\sum_{m=0}^{\infty} d_mm\mathcal{J}e^{-\frac{m(\mathcal{J}+\omega)}{T}},
\end{equation}
where $m$ is the highest weight of a particular representation, with degeneracy $d_m$ and energy $m\mathcal{J}$. Examples of non-singlet states are the $N^2$ number of 
$a^{\dagger i}_j|0\rangle$'s
in the adjoint representation with energy $\mathcal{J}$, and the $N^4$ number of $a^{\dagger i}_ja^{\dagger k}_l|0\rangle$'s in the tensor product representation with energy $2\mathcal{J}$.

In the oscillator model, we expect $d_m$ to scale as $d_m\sim N^{2m}$. This factor cancels with the factor $e^{\frac{-m \mathcal{J}}{T}}$, since  $\mathcal{J}>2T \log{N}$. Therefore, we estimate the energy expectation value to be  
\begin{equation}
\begin{split}
    \Tr(\rho_TH_G)<&\sum_m m\mathcal{J}e^{\frac{-m\omega}{T}}.\\
\end{split}
\end{equation}

This is an $O(1)$ number times the energy scale $\mathcal{J}$. So we conclude that, for a thermal state with temperature $T<\frac{\mathcal{J}}{2\log{N}}$, the energy expectation value of $H_G$ per oscillator is $O(1)$ times $\frac{\mathcal{J}}{N^2}$.
 \end{document}